\documentclass[aps,prb,twocolumn,floatfix,showpacs,superscriptaddress]{revtex4-2}
\usepackage{graphics}
\usepackage{epsfig}
\usepackage{times}
\usepackage{bm}
\usepackage{braket}
\usepackage{color}

\usepackage{amsmath}	
\begin{document}
\title{Spin-$S$ impurities with XXZ anisotropy in a spin-1/2 Heisenberg chain}
\author{Ayushi Singhania}
\email{Present address: Institute for Theoretical Solid State Physics, IFW Dresden, 01069 Dresden, Germany}
\affiliation{Department of Physical sciences, Indian Institute of Science Education and Research Mohali, Sector 81, S.A.S. Nagar, Manauli PO 140306, India}
\author{Masahiro Kadosawa}
\affiliation{Department of Physics, Chiba University, Chiba 263-8522, Japan}
\author{Yukinori Ohta}
\affiliation{Department of Physics, Chiba University, Chiba 263-8522, Japan}
\author{Sanjeev Kumar}
\affiliation{Department of Physical sciences, Indian Institute of Science Education and Research Mohali, Sector 81, S.A.S. Nagar, Manauli PO 140306, India}
\author{Satoshi Nishimoto}
\affiliation{Department of Physics, Technical University Dresden, 01069 Dresden, Germany}
\affiliation{Institute for Theoretical Solid State Physics, IFW Dresden, 01069 Dresden, Germany}

\date{\today}

\begin{abstract}
We study the effect of anisotropy of spin-$S$ impurities on an antiferromagnetic
SU(2) Heisenberg chain. The magnetic impurities are assumed to have
an XXZ-type anisotropic exchange with its neighboring sites. First, using
density-matrix renormalization group technique. we examine spin-spin correlation
function, instability of N\'eel order, and local spin susceptibility in the presence
of a single spin-$S$ impurity. Based on the results, we find that the types
of spin-$S$ impurities are classified into two groups:
$\rm(\hspace{.18em}i\hspace{.18em})$ nonmagnetic and $S=1$ impurities
enhance only short-range antiferromagnetic correlation and
$\rm(\hspace{.08em}ii\hspace{.08em})$ $S=1/2$ and $S>1$ impurities can,
in contrast, stabilize a long-range N\'eel order in the disordered SU(2) Heisenberg
chain. Then, we focus on the case of $S=1/2$ impurity as a representative of
$\rm(\hspace{.08em}ii\hspace{.08em})$ and investigate the evolution of
some experimentally observable quantities such as magnetization, specific
heat, and magnetic susceptibility, as a function of concentration and XXZ
anisotropy strength of the impurity. We confirm that the N\'eel order is
induced in the bulk spin chain in the presence of finite amount of easy-axis
XXZ $S=1/2$ impurities.
Furthermore, we recover some of the aforementioned features using
cluster mean-field theory, which allows us to present results on experimentally
accessible quantities at finite temperatures.
Interestingly, in the presence of uniform magnetic field,
the total magnetization exhibits a pseudo-gap behavior for low values of applied field.
We also discuss the dependence of NMR spectrum on various XXZ impurities
and identify that the spin state of Co impurity in SrCu$_{0.99}$Co$_{0.01}$O$_2$
is $S = 3/2$.
\end{abstract}

\maketitle

\section{Introduction}

It is well known that the presence of impurities in solids can lead to quantitative
changes in their properties. For example, a disordered metal is expected
to have higher resistivity compared to a defect-free metal~\cite{Dugdale2005},
transition temperature to superconducting order can be altered by the presence
of impurities~\cite{Maeda1990}, etc. However, in some cases, the presence of
impurities can even modify the qualitative behavior of the system. For instance,
metals can turn into insulators due to disorder-induced phenomenon known
as Anderson localization~\cite{Anderson1958, Ying2016}. Such a qualitative change
of behavior can also occur in magnets. A famous example is the disorder-induced
change in the order of phase transitions~\cite{Imry1979}. Defects in magnetic
materials can modify not only the ground-state properties but also the excitation
spectrum~\cite{Hammerath2011}. Substitution of a magnetic ion by a different ion
with the same or different spin, or a magnetic ion coupled to random spin in a lattice,
corresponds to the presence of defects~\cite{Eggert1992}. Low-dimensional systems
are very sensitive to disorder and often display dramatic effects in the presence of
impurities due to interplay between quantum effects, strong correlations, and
disorder~\cite{Anfuso2006}.

Some observations of impurity-induced effects in spin systems are emergence of
$S=1/2$ degrees of freedom at the edges, which occurs when Cu is doped
in a Haldane material~\cite{Hagiwara1990}. A low concentration of nonmagnetic
impurities induce a long-range magnetic order in spin-Peierls material
CuGeO$_3$~\cite{Hase1993, Grenier1997}. Similar observations have been made
for a two-leg spin-1/2 ladder compound SrCu$_2$O$_3$, where doping of
as low as $1\%$ Zn ($S_{\rm imp}=0$) results in antiferromagnetic (AFM) 
behavior; we denote the magnitude of impurity spin by
$S_{\rm imp}$ and a nonmagnetic impurity is expressed as $S_{\rm imp}=0$.
Furthermore, it was shown that corresponding N\'eel temperature
can be increased with increase in concentration of impurities~\cite{Azuma1997}.
Several experimental studies have been performed on spin-1/2 materials 
Sr$_2$CuO$_3$ and SrCuO$_2$, which are considered good realizations of
one-dimensional (1D) Heisenberg model~\cite{Motoyama1996}.

Experimental investigations, on the other hand,  reveal an enhancement of
long-range order on the introduction of Zn$^{2+}$ ($S_{\rm imp}=0$) in
SrCuO$_2$ (zigzag chain) and Sr$_2$CuO$_3$ (linear chain)~\cite{Kojima2004}.
Whereas, substitution of $S_{\rm imp}=1$ impurity in a spin-1/2 Heisenberg chain
is known to result in a Kondo-singlet where the impurity spin is Kondo screened by
the two neighboring spins of the chain. Similar to nonmagnetic substitution, formation
of singlets at $S_{\rm imp}=1$ impurity site disrupts the translational invariance
of the chain, breaking it into finite lengths. This leads to confinement of
spinons and results in emergence of a spin gap in low-lying excitations. This
has been confirmed experimentally for low concentration of Ni ($S_{\rm imp}=1$)
doping in SrCuO$_2$, where sizeable spin pseudogap appears as a consequence
of impurities~\cite{Simutis2013}. While experimental results reveal that doping of
$S_{\rm imp}=0, 1$ in spin chain materials suppresses long-range magnetic ordering temperature~\cite{Karmakar2017}. Investigations of replacing a spin-1/2
magnetic ion (Cu$^{2+}$) with another spin-1/2 ion (Co$^{2+}$) in spin chain
material SrCuO$_2$ reveal that the bulk behavior switches from Heisenberg to
Ising-like. Due to this, Ising-like (or XXZ-type) anisotropy-induced  magnetic
ordering temperature is enhanced; however gapless nature of the spin excitations
are not disturbed. Similar behaviors of N\'eel-type ordering appear in
Co-based spin-1/2 Ising chain compounds BaCo$_2$V$_2$O$_8$~\cite{Kimura2008}
and SrCo$_2$V$_2$O$_8$~\cite{Bera2014}. However, in the Co-doped SrCuO$_2$,
the spin-lattice relaxation rates $T_1^{-1}$ does not obey the gaplike decrease,
and the possibility of Co$^{2+}$ acting as a spin-3/2 ion has also been
suggested~\cite{Utz2017}.

Theoretical efforts using field theory, renormalization
arguments~\cite{Eggert1992, Eggert2001}, and numerical methods like quantum
Monte Carlo~\cite{Bobroff2009, Alexander2010} have been successfully
employed to investigate properties of low-dimensional materials when
doped with magnetic (specifically $S_{\rm imp}=1$) and nonmagnetic impurities.
Field theoretical and numerical studies using density-matrix renormalization
group (DMRG) of isotropic spin-1/2 impurity coupled to spin-1/2 chain expects
the impurity spin to be over-screened in analogy to Kondo
effect~\cite{Eggert2001,Zhang1997, Zhang1997-2}. However, experimental
results with $S_{\rm imp}=1/2$ embedded in the chain stress the importance
of anisotropic effects. Also, very few studies have been carried out on
the effect of magnetic impurity with $S_{\rm imp}>1$.

In this paper, motivated by the above situation, we consider the spin-1/2
AFM Heisenberg chain doped with XXZ-type anisotropic
spin-$S$ impurities using DMRG technique and cluster mean-field theory (CMFT)
approach. Based on the ground-state properties in the presence of a single
spin-$S$ impurity, we find that the types of spin-$S$ impurities are classified
into two groups:
$\rm(\hspace{.18em}i\hspace{.18em})$ nonmagntic and $S=1$ impurities
enhance only short-range antiferromagnetic correlation and
$\rm(\hspace{.08em}ii\hspace{.08em})$ $S=1/2$ and $S>1$ impurities can,
in contrast, stabilize a long-range N\'eel order in the disordered SU(2) Heisenberg
chain. Then, we focus on the case of $S=1/2$ impurity as a representative of
$\rm(\hspace{.08em}ii\hspace{.08em})$ in the latter part of this paper.
We thus confirm that the N\'eel order is induced in the bulk spin chain in
the presence of finite amount of easy-axis XXZ $S=1/2$ impurities and
the corresponding staggered magnetization increases with increasing the
impurity density as well as the anisotropy strength. Interestingly, in presence
of uniform magnetic field, total magnetization exhibits a pseudo-gap behavior
for low values of applied field. We also show that some of
the above features can be obtained using CMFT. This allows us to discuss
the finite temperature behavior of the model in the context of real materials,
where the spin chains are typically weakly coupled to each other and
finite-temperature phase transitions are possible.

The remainder of the paper is organized as follows:  In Sec.~II, we explain the spin-1/2
AFM Heisenberg model doped with XXZ-type anisotropic spin-$S$ impurities
and describe the numerical methods applied. In Sec.~III, we examine the
effect of spin-$S$ XXZ impurities on the ground-state properties. In Sec.~IV,
the temperature dependence of specific heat and spin susceptibility
as functions of the impurity density and the anisotropy strength is discussed.
Section~V provides a summary and conclusions.

\section{Model and method}

\subsection{Impurity-doped spin-1/2 Heisenberg chain}

We consider a spin-1/2 AFM Heisenberg chain and replace a finite number of
sites with spin-$S$ impurities. The impurities are assumed to have an XXZ-type
easy-axis anisotropy on the interaction links with its neighbors. The Hamiltonian reads
\begin{align}
\nonumber
	{\cal H} =& \sum_{i \notin {\rm imp.}} [\frac{1}{2}(S^+_iS^-_{i+1}+S^-_iS^+_{i+1})+ S^z_iS^z_{i+1}]\\
\nonumber
		 &+\sum_{j \in {\rm imp.}} [\frac{1}{2}(S^-_{j-1}+S^-_{j+1}) S^+_{{\rm imp},j}+{\rm H.c.} \\
		 &\ \ \ \ \ \ \ \ \ \ \ \ \ \ +\Delta_{\rm imp} (S^z_{j-1}+S^z_{j+1}) S^z_{{\rm imp},,j}],
	\label{eq:hamiltonian}
\end{align}
where $\mathbf{S}_i$ is spin-$\frac{1}{2}$ operator at non-impurity site $i$,
$\mathbf{S}_{{\rm imp},,j}$ is spin-$S$ operator at impurity site $j$, and
$\Delta_{\rm imp}$ is the XXZ anisotropy of the interaction between the impurity
and neighboring sites. We assume that the undoped chain is spin-isotropic.
The impurity density $n_{\rm imp}$ is defined as the ratio of 
$N_{\rm imp}/N_c$, where $N_{\rm imp}$ is number of impurities embedded
in the chain consisting of $N_c$ spins.
Note that, in the limit of $n_{\rm imp}=1$ for $S_{\rm imp}=1/2$,
the Hamiltonian is reduced to a spin-1/2 Heisenberg chain with the uniform XXZ
anisotropy $\Delta_{\rm imp}$.

\begin{figure}[tbh]
\centering
\includegraphics[width=1.0\linewidth]{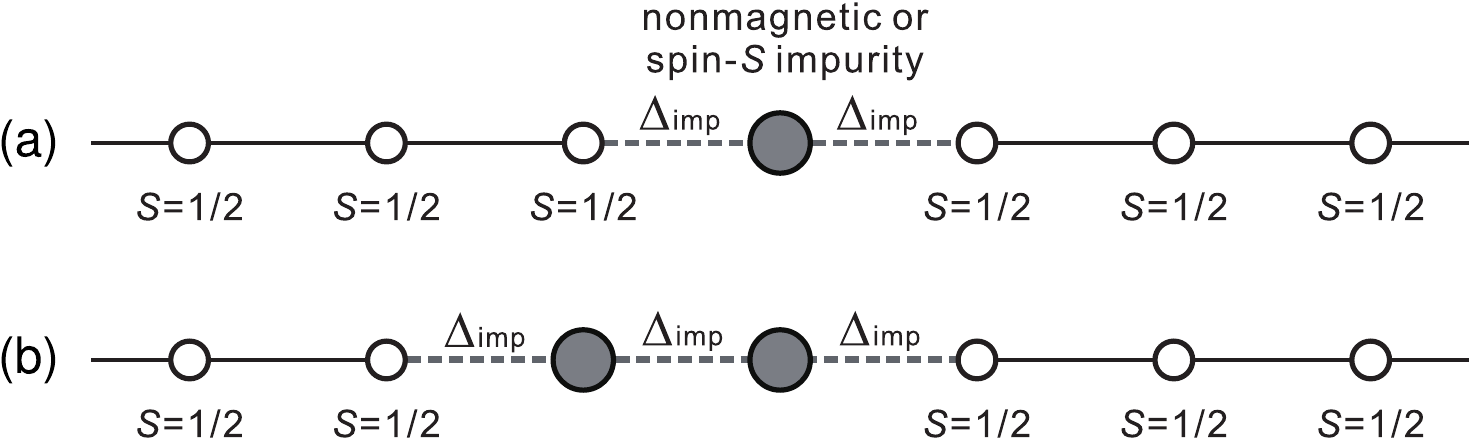}
\caption{
(a) A schematic view of spin-1/2 Heisenberg chain with an internal nonmagnetic
(vacancy) or spin-$S$ impurity (filled circle). The exchange interaction on the
undoped chain is spin-isotropic (solid line) and that between the magnetic impurity
and its neighboring sites can be of the XXZ-type. In (b), two impurities sit on adjacent sites. 
}
\label{lattice}
\end{figure}

For a given set of $N_{\rm imp}$ and $N_c$, the positions of impurities are
randomly distributed. Then, setting the length of non-impurity chain between
two impurities to be $l$, the distribution function of existing probability is
given by
\begin{align}
	P(l)=n_{\rm imp}\exp\left[-\frac{n_{\rm imp}}{\sqrt{1-n_{\rm imp}}}l\right].
	\label{probability}
\end{align}
The mean length of non-impurity chain is $\bar{l}=(1-n_{\rm imp})/n_{\rm imp}$.

\subsection{Density-matrix renormalization group}

To examine the ground-state properties of impurity-doped chain, we employ
the DMRG technique, which is a powerful numerical method for various
1D quantum systems~\cite{White92}. Although the DMRG is basically
restricted to the ground-state calculations, very long chains with order
of a few thousand sites can be studied with high accuracy. Hence, a good
realization of randomly distributed impurities close to the bulk limit is possible.
We keep up to $m=6000$ density-matrix eigenstates in the renormalization
procedure. In this way, the discarded weight is less than $1 \times10^{-13}$.
However, the DMRG wave function frequently tends to get trapped in
a “false” (or metastable) ground state when we study disordered systems.
Thus, we need to pay special attention to the convergence of calculation, 
for example, by such actions as confirming the unchanged convergence
even with different initial conditions. Either open or periodic boundary
condition is chosen, depending on the calculated quantities.

\subsection{Cluster mean-field technique}

We also utilize CMFT to study the Hamiltonian with spin-1/2 XXZ impurities
on 1D isotropic SU(2) cluster. The CMFT is an extension of a single-site
Weiss mean-field (MF) theory, where instead of a single site we consider
a cluster comprising of $N_c$ spins. For 1D system, the edge spins
$\mathbf{S}_1$ and $\mathbf{S}_{N_c}$ couple to neighboring cluster via
standard mean-field decoupling. Since, there are numerous ways in which
impurity sites can be distributed, we average observables over various random
configurations $N_{av}$ for a fixed number of impure sites. The edge spins
are forced to remain pure in a random configuration to avoid explicit dependence
of anisotropy due to the MF decoupling. As a result, the impurity density, 
$n_{\rm imp}$, can reach a maximum of $(N_c-2)/N_c$. We present results
computed for a cluster of $10$ spins (unless specified otherwise). Calculations
are also performed for larger cluster sizes for scaling analysis.

The CMFT enables us to study phase transition in thermodynamic properties
within a given cluster size. Although the Mermin-Wagner theorem rigorously
forbids finite-temperature phase transition associated with a spontaneous
symmetry breaking in pure 1D systems due to thermal and quantum fluctuations,
we may argue that any symmetry-breaking MF-type treatment of quantum systems
should provide a finite-temperature phase transition. However,
this becomes physically relevant for real systems where the 1D spin chains
are typically weakly coupled with neighboring chains and finite transition
temperatures are experimentally observed. The transition temperatures
may well be low if partial inclusion of quantum fluctuations is achieved by, e.g.,
the CMFT as in the present case. Most importantly, the dependence
of transition temperatures on the strength of anisotropy and concentration
of impurities as obtained via CMFT can provide direct experimentally
measurable consequences of such impurity doping.

\section{Ground-state properties}

\subsection{Influence of single impurity on spin-spin correlations}

\begin{figure}[tbh]
\centering
\includegraphics[width=1.0\linewidth]{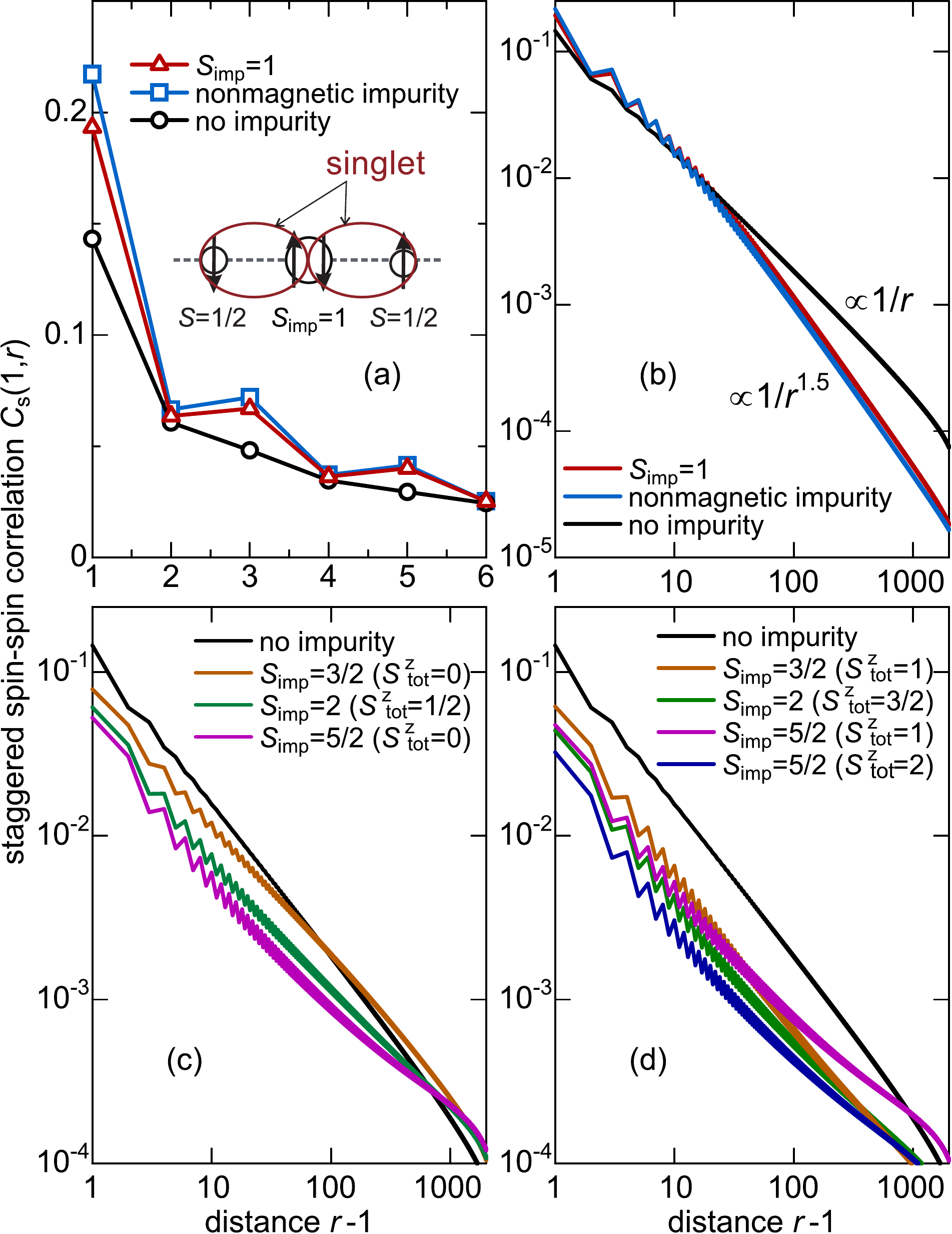}
\caption{
(a,b) DMRG results for the staggered spin-spin correlation $C_{\rm s}(1,r)$ with
nonmagnetic impurity and $C_{\rm s}(2,r)$ with $S_{\rm imp}=1$ impurity.
Inset in (a): Structure of two spin-singlet pairs around the $S_{\rm imp}=1$ impurity.
(c,d) Log-log plots of $C_{\rm s}(1,r)$ with $S_{\rm imp}>1$ impurities. 
In each plot the spin-spin correlation of the bulk SU(2) Heisenberg
chain is shown as a reference.
}
\label{fig_stagszsz}
\end{figure}

We begin by investigating the effects of single XXZ spin-$S$
impurity on a spin-1/2 Heisenberg chain -- a good representation of very low
concentration of magnetic impurities in a real material. To do so, we first
calculate the distance dependence of spin-spin correlation functions from the impurity.
Since an AFM Heisenberg chain is considered, we need to focus mainly on
the staggered spin-spin correlation function, which is defined as
\begin{align}
C_{\rm s}(i,j)=(-1)^{|i-j|}\langle S^z_i S^z_j \rangle,
	\label{eq:spincorr}
\end{align}
where $\langle {\cal O} \rangle$ denotes expectation value of operator ${\cal O}$
in the ground state. A site next to impurity is indexed by `1'.

\subsubsection{Nonmagnetic impurity}

In the case of nonmagnetic impurity ($S_{\rm imp}=0$), i.e, vacancy, an enhancement
of short-range staggered spin-spin correlations near the impurity has been theoretically suggested~\cite{Eggert95,Laukamp98}. We here confirm it and also see the long-range
spin-spin correlation from the impurity. Since the original spin chain is simply cut by
a nonmagnetic impurity, no correlations exist between spins on both sides of the
nonmagnetic impurity. Therefore, we only need to study an open chain. In
Fig.~\ref{fig_stagszsz}(a), the short-range behavior of staggered spin-spin correlation
function $C_{\rm s}(1,r)$ for $S_{\rm imp}=0$ is compared to that of a bulk SU(2)
Heisenberg chain without impurity.  We see that the AFM correlation
between site at $1$ and $2$ is significantly increased from $C_{\rm s}(1,2)=0.1456$
for the bulk chain to $C_{\rm s}(1,2)=0.2172$ for $S_{\rm imp}=0$ impurity.
These values agree perfectly to those from the Bethe-ansatz
analysis~\cite{Frahm97}. Furthermore, such an enhancement of $C_{\rm s}(1,r)$
is seen at certain distances ($r\lesssim10$) especially for even $r$. This is consistent
with the previous numerical study~\cite{Laukamp98}. Then, we explore what
happens in the correlation at large distances. Figure~\ref{fig_stagszsz}(b) shows log-log
plots of $C_{\rm s}(1,r)$ as a function of distance from the $S_{\rm imp}=0$ impurity. 
It is confirmed that the staggered oscillations are maintained indefinitely because
$C_{\rm s}(1,r)$ exhibits an asymptotic behavior with keeping its positive value.
The decay rate is $\propto r^{-1.5}$. On the other hand, as is well known,
the spin-spin correlation of the bulk SU(2) Heisenberg chain decays as
$\propto r^{-1}$ [see Fig.~\ref{fig_stagszsz}(b)]. The decay of $C_{\rm s}(1,r)$
for nonmagnetic impurity is obviously faster than that for the bulk chain. In short,
the short-range staggered spin-spin correlation is indeed enhanced by nonmagnetic
impurity but the long-range correlation is rather suppressed.

\subsubsection{Magnetic impurity with $\Delta_{\rm imp}=1$}

We then examine the effect of magnetic impurity. 
Unlike the case of nonmagnetic impurity, the spins on both sides of
magnetic impurity are still correlated. Hence, only the replacement of
open-chain ends by magnetic impurities is not adequate to evaluate
the effect of the magnetic impurity. Therefore, we shall proceed as follows:
We prepare an open long SU(2) Heisenberg chain with length $2000-4000$
and replace the $500$th site from one end by a magnetic impurity; and then,
calculate the correlation $C_{\rm s}(1,r+1)$ with setting the $501$th site
as $i=1$, where the distance $r$ is counted towards another chain end.
In this way, the Friedel oscillations arising from the open ends can be
negligibly small around the impurity at $500$th site. We have also confirmed
that the results are unchanged even if the position of magnetic impurity is
shifted by a few sites from $500$th site. First, we restrict ourselves to the
case of $\Delta_{\rm imp}=1$.

Interestingly, the results for $S_{\rm imp}=1$ impurity are even quantitatively
similar to those for nonmagnetic impurity. As seen in Fig.~\ref{fig_stagszsz}(a),
the staggered spin-spin correlation is significantly enhanced around the $S_{\rm imp}=1$
impurity , e.g., $C_{\rm s}(2,3)=0.1933$ in contrast to $C_{\rm s}(1,2)=0.1456$
for the bulk chain. Note that we look at $C_{\rm s}(2,j)$ instead of $C_{\rm s}(1,j)$
in the $S_{\rm imp}=1$ case. This can be explained as follows: The $S_{\rm imp}=1$
impurity is fractionalized into two spin-1/2's and each of them forms a spin singlet with
the neighboring spin-1/2 site. Accordingly, the $S_{\rm imp}=1$ impurity and the
neighboring two spin-1/2's are screened, and the three sites may behave like
a nonmagnetic impurity [see the inset of Fig.~\ref{fig_stagszsz}(a)]. Therefore,
in the physical sense it is more reasonable to see not $C_{\rm s}(1,j)$ but
$C_{\rm s}(2,j)$ in the $S_{\rm imp}=1$ case. Thus, we find that the values
of $C_{\rm s}(2,j)$ for $S_{\rm imp}=1$ impurity are almost equivalent to those of
$C_{\rm s}(1,j)$ for nonmagnetic impurity at short distance. Furthermore,
it is surprising that the asymptotic behavior of $C_{\rm s}(2,j)$ for
$S_{\rm imp}=1$ impurity is $\propto r^{-1.5}$ as that of $C_{\rm s}(1,j)$
for nonmagnetic impurity [see Fig.~\ref{fig_stagszsz}(b)].
It seems that the influence of nonmagnetic and $S_{\rm imp}=1$ impurities
on the spin-spin correlations is almost identical even in the quantitative sense,
although it would be a natural consequence of the fact that an $S_{\rm imp}=1$
impurity behaves like a nonmagnetic impurity. This also implies that
the valence bond formations of the $S_{\rm imp}=1$ impurity and its neighboring
sites is quite robust.

Moreover, the staggered spin-spin correlation in the presence of higher-$S$
magnetic impurities ($S_{\rm imp}>1$) is investigated. In Fig.~\ref{fig_stagszsz}(c,d),
we plot DMRG results for $C_{\rm s}(1,j)$ with $S_{\rm imp}>1$ impurities,
where the correlation of the bulk SU(2) Heisenberg chain is also shown as
a reference data. Note that, when the SU(2) Heisenberg chain is doped
with a spin-isotropic ($\Delta_{\rm imp}=1$) $S_{\rm imp}>1$ impurity,
the ground state is degenerate in the $S^z_{\rm tot}=0$, $1$, $\cdots$,
$S_{\rm imp}-1/2$ sectors for odd $S_{\rm imp}$ and in the $S^z_{\rm tot}=1/2$,
$3/2$, $\cdots$, $S_{\rm imp}-1/2$ sectors for even $S_{\rm imp}$. 
Accordingly, the results for the ground state in the lowest $S^z_{\rm tot}$
sector for each $S_{\rm imp}$ impurity are plotted in Fig.~\ref{fig_stagszsz}(c)
and those in the higher $S^z_{\rm tot}$ sectors are plotted in Fig.~\ref{fig_stagszsz}(d).
We find that the short-range staggered spin-spin correlation is decreased with
increasing the magnitude of $S_{\rm imp}$. It may be naturally expected
because the AFM fluctuations around higher $S_{\rm imp}$ impurity are
suppressed due to weaker quantum fluctuations closer to the classical limit.
However, the decay rate of spin-spin correlation for any $S_{\rm imp}$ impurity
seems to comparable to that of the bulk chain, i.e., $\propto r^{-1}$ at
short distance ($r \lesssim 100$). We then see an uncommon behavior at
larger distance; $C_{\rm s}(1,j)$ approaches that of the bulk chain at
intermediate distance ($100 \lesssim r \lesssim 1000$) and returns to exhibit
$C_{\rm s}(1,j) \propto r^{-1}$ again at large distance ($r \gtrsim 1000$).
Although we have confirmed that this uncommon behavior is not an artifact
due to finite-size effects by studying several chains with lengths
$L=2000-4000$,   the reason why that is so is currently unclear.

In summary, the influence of single impurity on spin-spin correlation is classified
broadly into two kinds: (i) $S_{\rm imp}=0$, $1$ with an enhancement of
short-range staggered correlation and its decay as $C_{\rm s}(1,r+1)\propto r^{-1.5}$
and (ii) $S_{\rm imp}>1$ with a suppression of short-range staggered correlation
and its decay as $C_{\rm s}(1,r+1)\propto r^{-1}$.

\begin{figure}[tbh]
\centering
\includegraphics[width=1.0\linewidth]{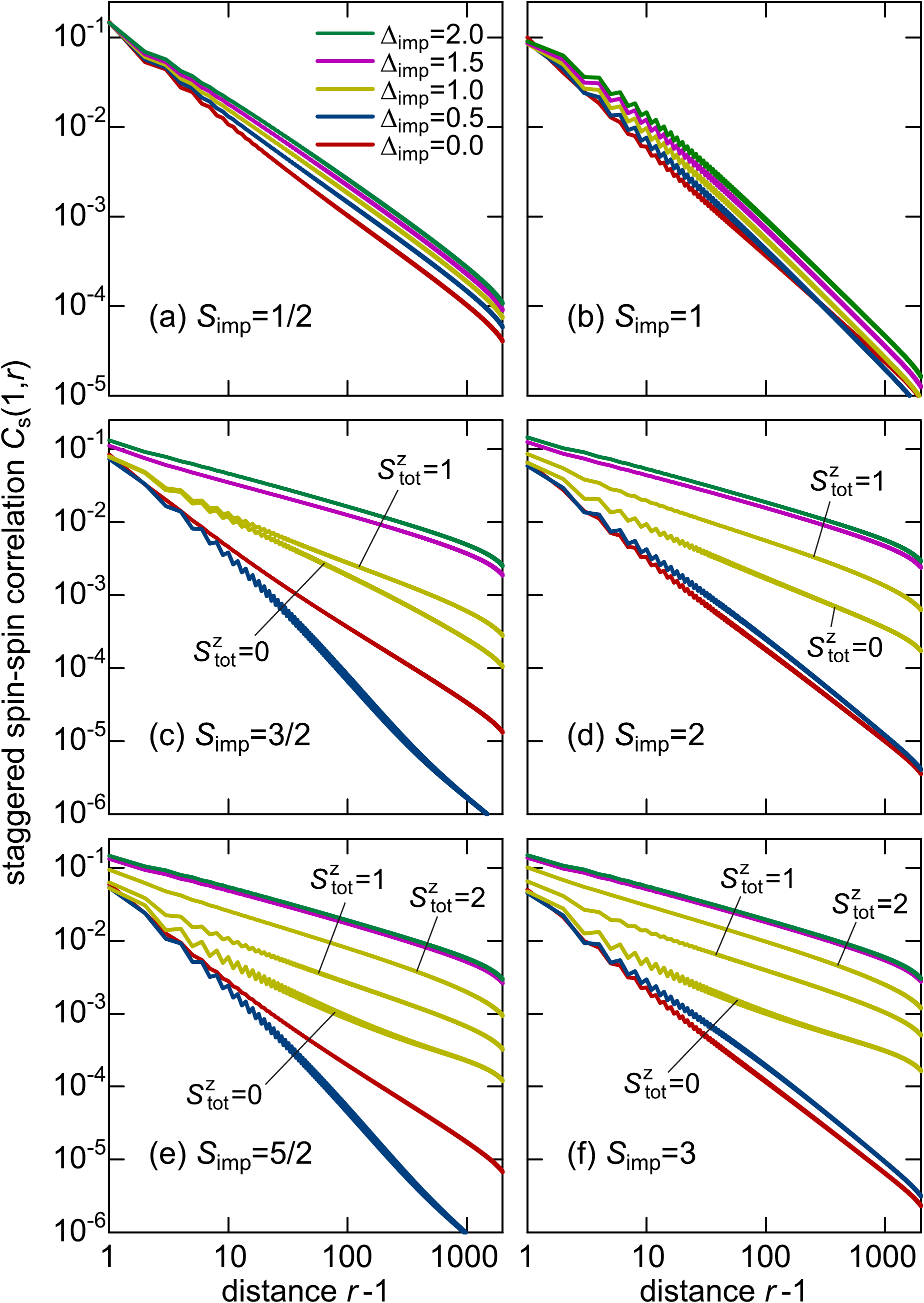}
\caption{
Log-log plots of the staggered spin-spin correlation function $C_{\rm s}(1,r)$
for (a) $S_{\rm imp}=1/2$, (b) $S_{\rm imp}=1$, (c) $S_{\rm imp}=3/2$,
(d) $S_{\rm imp}=2$, (e) $S_{\rm imp}=5/2$, and (f) $S_{\rm imp}=3$
impurities at several $\Delta_{\rm imp}$ values. Note that the case of
$\Delta_{\rm imp}=1.0$ in (a) is equivalent to the balk chain.
}
\label{fig:szsz_delta}
\end{figure}

\subsubsection{Magnetic impurity with $\Delta_{\rm imp}\neq1$}

Next, we examine how the staggered spin-spin correlation is affected by the
XXZ anisotropy ($\Delta_{\rm imp}$) of exchange interaction at magnetic impurity.
Figure~\ref{fig:szsz_delta}(a) shows DMRG result for $C_{\rm s}(1,r)$ at
several values of $\Delta_{\rm imp}$ with $S_{\rm imp}=1/2$ impurity, 
where the result at $\Delta_{\rm imp}=1$ is equivalent to that of the balk SU(2) 
Heisenberg chain. Since, as discussed below in Sec.~\ref{sec_neel}, the AFM
fluctuations around the impurity are sensitive to $\Delta_{\rm imp}$, one may expect
a large dependence of $C_{\rm s}(1,r)$ on $\Delta_{\rm imp}$. However, in fact,
the sort-range $C_{\rm s}(1,r)$ is hardly affected by $\Delta_{\rm imp}$ and
the asymptotic behavior is only slightly changed as $C_{\rm s}(1,j)\propto r^{-1.05}$ at
$\Delta_{\rm imp}=0$ and $C_{\rm s}(1,j)\propto r^{-0.95}$ at $\Delta_{\rm imp}=2$.
As shown in Fig.~\ref{fig:szsz_delta}(b), DMRG results for the
$\Delta_{\rm imp}$-dependence of $C_{\rm s}(2,r)$ with the $S_{\rm imp}=1$
impurity exhibit qualitatively similar behaviors to  those in the case of $S_{\rm imp}=1/2$
impurity. Although the short-range $C_{\rm s}(2,r)$ is only slightly larger for larger
$\Delta_{\rm imp}$, the decay rate at large distance is almost independent of
$\Delta_{\rm imp}$, i.e., $C_{\rm s}(2,r) \propto r^{-1.5}$.
This implies that an $S_{\rm imp}=1$ impurity behaves like a nonmagnetic
impurity for any $\Delta_{\rm imp}$. This is another evidence to prove that 
the valence-bond picture depicted in the inset of Fig.~\ref{fig_stagszsz}(a) is still
a good approximation.

Interestingly, as shown in Fig.~\ref{fig:szsz_delta}(c-f), the $\Delta_{\rm imp}$
dependence of staggered spin-spin correlation functions $C_{\rm s}(1,r)$
look similar for any $S_{\rm imp}>1$ impurities. Their qualitative trends are
also similar to those for the $S_{\rm imp}=1/2$ impurity; however, the effect
of $\Delta_{\rm imp}$ seems to be much more pronounced. Basically, the decay
of $C_{\rm s}(1,r)$ as a function of $r$ is faster (slower) for smaller (larger)
$\Delta_{\rm imp}$.The only exception is that there seems to exist a finite
$\Delta_{\rm imp}$ which gives the fastest decay of $C_{\rm s}(1,r)$ for
the $S_{\rm imp}=3/2$ and $5/2$ impurities. Note that, for $S_{\rm imp}>1$
impurities, the DMRG results for all $S^z_{\rm tot}$ sectors giving degenerate
ground states are shown.

\subsection{Instability of N\'eel order with impurities}~\label{sec_neel}

\begin{figure}[t]
\centering
\includegraphics[width=1.0\linewidth,clip]{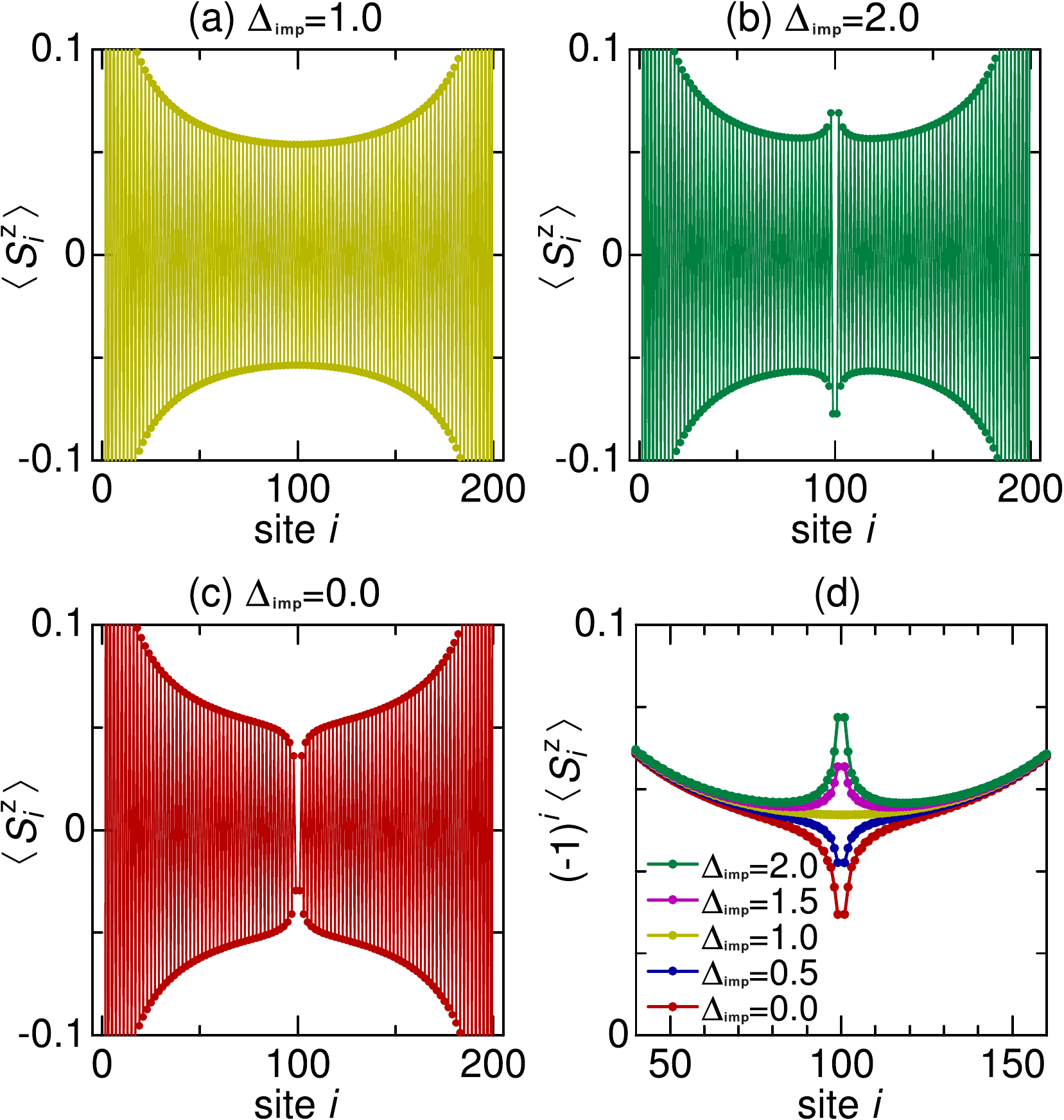}
\caption{
(a-c) Expectation values of the $z$-component of local spin, $\langle S^z_i \rangle$,
in a 200-site SU(2) Heisenberg open chain with an XXZ impurity at the center.
(d) $\Delta_{\rm imp}$-dependence of $(-1)^i\langle S^z_i \rangle$ around the XXZ impurity.
}
\label{fig:dejtadepS12}
\end{figure}

In the previous subsection, the development of  staggered spin-spin correlation
on one side of an XXZ magnetic impurity is discussed. Nothing about correlation
between spins on both sides of the XXZ magnetic impurity has so far been considered. 
For example, we know that a nonmagnetic impurity enhances the staggered
oscillations on each side of the impurity. However, since the nonmagnetic impurity
completely cuts the original SU(2) Heisenberg chain, no correlations exist between
the separated chains. One could interpret this to mean that the intrinsic
quasi-long-range N\'eel order in the SU(2) Heisenberg chain is broken by the doping
of nonmagnetic impurity. Therefore, we may say that nonmagnetic impurity gives
a negative contribution to the stabilization of long-range N\'eel order.

We then investigate how the quasi-long-range N\'eel order of the SU(2)
Heisenberg chain is developed by an XXZ magnetic impurity.
We start with an $S_{\rm imp}=1/2$ impurity, where the case of
$\Delta_{\rm imp}=1$ corresponds to no impurity. As a reference state,
we prepare quasi staggered order in a 200-site SU(2) Heisenberg open chain,
where the $z$-component of edge spins are fixed at $\langle S^z_1 \rangle=-1/2$
and $\langle S^z_{200} \rangle=1/2$ in order to control the staggered phase
of $\langle S^z_i \rangle$. The spatial distribution of $\langle S^z_i \rangle$
in the reference state, i.e., without impurity, is shown in Fig.~\ref{fig:dejtadepS12}(a).
The staggered oscillation of $\langle S^z_i \rangle$ is created as
a consequence of the Friedel oscillation from both chain ends.
Despite the decay of $|\langle S^z_i \rangle|$ with $\sim 1/r$ near
the both ends~\cite{Affleck1989,Singh1989}, the amplitude of
$|\langle S^z_i \rangle|$ is almost uniform around the center
of 200-site chain so that the effect of XXZ magnetic impurity can be clearly
demonstrated by replacing the central ($i=100$) site with the impurity.

It is generally known that the spin-1/2 XXZ Heisenberg chain has
a long-range N\'eel order with an easy-axis
anisotropy~\cite{Hulthen1938,Yang1966-1,Yang1966-2} and
a power-law decay of spin-spin correlation
$\langle S^z_i S^z_{i+r} \rangle \sim 1/r^\eta$ ($\eta>1$) with
an easy-plane anisotropy~\cite{Luther1975,Bogoliubov1986}. 
By analogy with this fact, one may naively expect that the
N\'eel stability is enhanced (suppressed) by an XXZ impurity with
$\Delta_{\rm imp}>1$ ($\Delta_{\rm imp}<1$).
In fact, this speculation is confirmed by the enhancement (suppression)
of amplitude of $\langle S^z_i \rangle$ around the XXZ magnetic impurity
with $\Delta_{\rm imp}=2$ ($\Delta_{\rm imp}=0$) as demonstrated in
Fig.~\ref{fig:dejtadepS12}(b,c). Accordingly, a finite doping of XXZ magnetic
impurities with $\Delta_{\rm imp}>1$ onto an SU(2) Heisenberg chain
stabilizes a long-range N\'eel order since the undoped chain is critical.
If we define staggered magnetization, as an order parameter of the
N\'eel state, by
\begin{align}
m_{\rm st}^z=\frac{1}{L}\sum_1^L |(-1)^i \langle S^z_i \rangle|,
	\label{eq:stm}
\end{align}
$m_{\rm st}^z>0$ should be achieved for any finite density of XXZ
impurities with $\Delta_{\rm imp}>1$ in the thermodynamic limit.
Details are discussed in Sec.~\ref{sec:mag}. Furthermore, interestingly,
the XXZ impurities with $\Delta_{\rm imp}>1$ would similarly
stabilize a long-range N\'eel order even if the undoped chain is an XXZ
Heisenberg chain with easy-plane anisotropy because of the power-law
decay of spin-spin correlation in the absence of impurity. To summarize
the effect of spin-1/2 XXZ impurity on $\langle S^z_i \rangle$,
we plot $(-1)^i \langle S^z_i \rangle$ as a function of $i$ for
several $\Delta_{\rm imp}$ values in Fig.~\ref{fig:dejtadepS12}(d).
We find that the increase or decrease of amplitude of $\langle S^z_i \rangle$
at the XXZ impurity is roughly proportional to $\Delta_{\rm imp}-1$.

\begin{figure}[tb]
\centering
\includegraphics[width=1.0\linewidth]{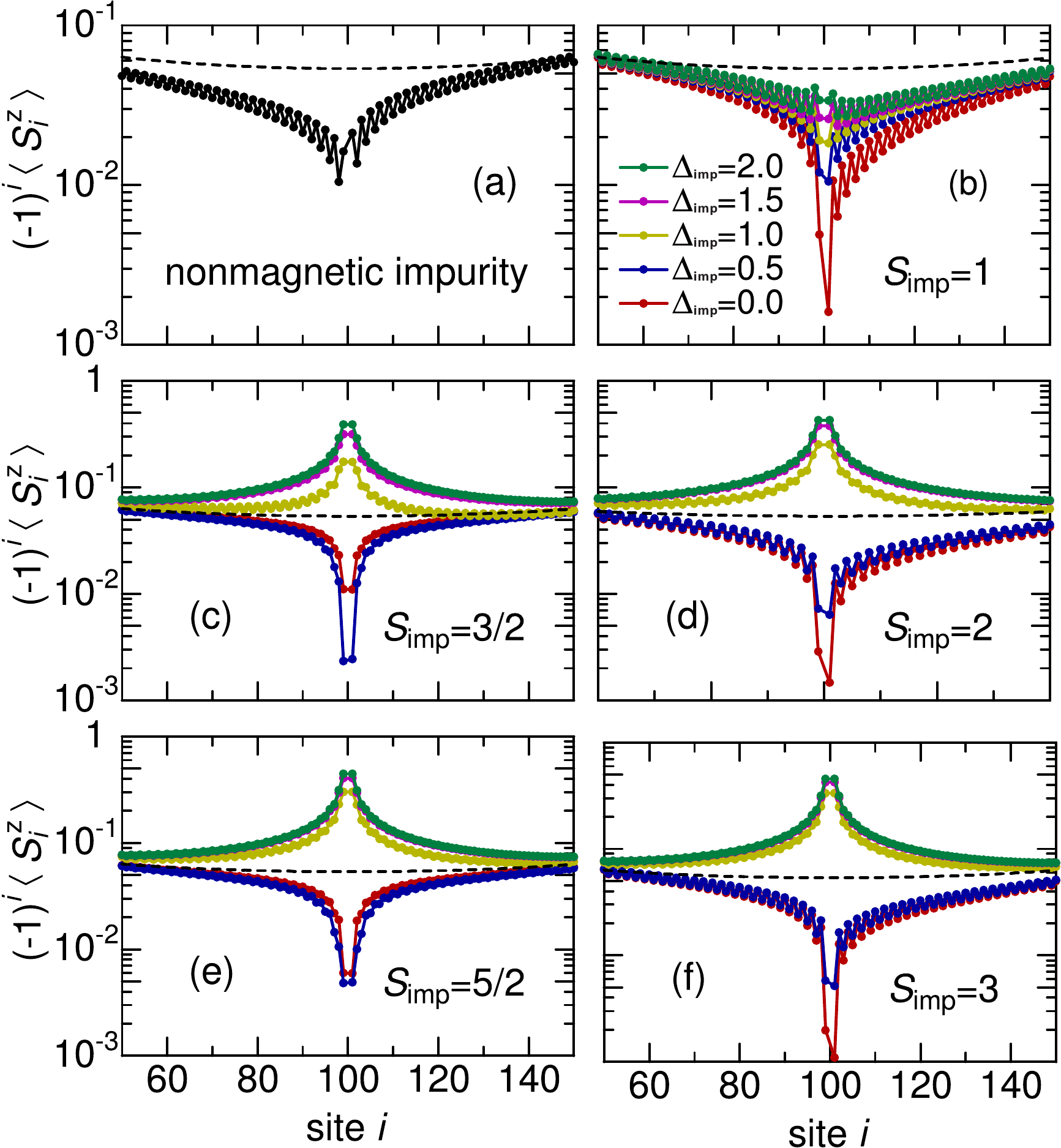}
\caption{
$\Delta_{\rm imp}$-dependence of staggered $z$-component of local spin
around the XXZ magnetic impurity for (a) $S_{\rm imp}=0$, (b) $S_{\rm imp}=1$,
(c) $S_{\rm imp}=3/2$, (d) $S_{\rm imp}=2$, (e) $S_{\rm imp}=5/2$,
and (f) $S_{\rm imp}=3$.
}
\label{fig:sz_sdep}
\end{figure}

So, what happens if an SU(2) Heisenberg chain is doped with XXZ impurity
other than $S_{\rm imp}=1/2$? To tell the conclusion first, the effects of
$S_{\rm imp}>1$ impurities are qualitatively similar to those in the
$S_{\rm imp}=1/2$ case, whereas the nonmagnetic and $S_{\rm imp}=1$
impurities lead to different behaviors. As done in the case of
$S_{\rm imp}=1/2$ impurity, we examine the change of $\langle S^z_i \rangle$
when the central site of 200-site quasi-long-range staggered order is
replaced by a nonmagnetic or an XXZ magnetic impurity. Fig.~\ref{fig:sz_sdep}(a)
shows DMRG result for $(-1)^i \langle S^z_i \rangle$ in the case of nonmagnetic
impurity. The original staggered oscillation is suppressed around the nonmagnetic
impurity because the chain is cut off by the impurity. As expected, 
a similar behavior is observed in the case of $S_{\rm imp}=1$ impurity.
As seen in Fig.~\ref{fig:sz_sdep}(b), the staggered oscillation is suppressed
around the $S_{\rm imp}=1$ impurity even with large easy-axis anisotropy
$\Delta=2$. This would be a natural consequence of the fact that 
an $S_{\rm imp}=1$ XXZ impurity for any $\Delta_{\rm imp}$ acts like
a nonmagnetic impurity, as discussed above. Therefore, we conclude that
nonmagnetic and $S_{\rm imp}=1$ impurities only give a negative contribution
for the stability of N\'eel order, namely, $m_{\rm st}^z=0$ is always obtained
when an SU(2) Heisenberg chain is doped with whatever amount of nonmagnetic
and $S_{\rm imp}=1$ impurities.

On the other hand, we find that the original staggered oscillation can be enhanced
by $S_{\rm imp}>1$ impurities if $\Delta_{\rm imp}$ is larger than a certain value
$\Delta_{{\rm imp},c}$. The results are shown in Fig.~\ref{fig:sz_sdep}(c-f). The overall
features are similar to those in the case of $S_{\rm imp}=1/2$ impurity. Nevertheless,
it is interesting to see that the staggered oscillation is enhanced by the $S_{\rm imp}>1$
impurities even with isotropic interaction $\Delta_{\rm imp}=1$ unlike in the case of
$S_{\rm imp}=1/2$ impurity. This may be because the higher $S_{\rm imp}$
impurity has a larger Ising anisotropy, due to its classical nature, than the
$S_{\rm imp}=1/2$ impurity. The critical values are estimated as
$\Delta_{{\rm imp},c}=0.918$, $0.997$, $0.994$, and $0.998$ for 
$S_{\rm imp}=3/2$, $2$, $5/2$, and $3$ impurities, respectively.
Thus, we confirm that a long-range N\'eel order can be stabilized once
an SU(2) Heisenberg chain is doped by $S_{\rm imp}>1$ impurities with
$\Delta \gtrsim 1$.

\subsection{Local spin susceptibility}~\label{sec:lss}

\begin{figure}[tbh]
\centering
\includegraphics[width=1.0\linewidth]{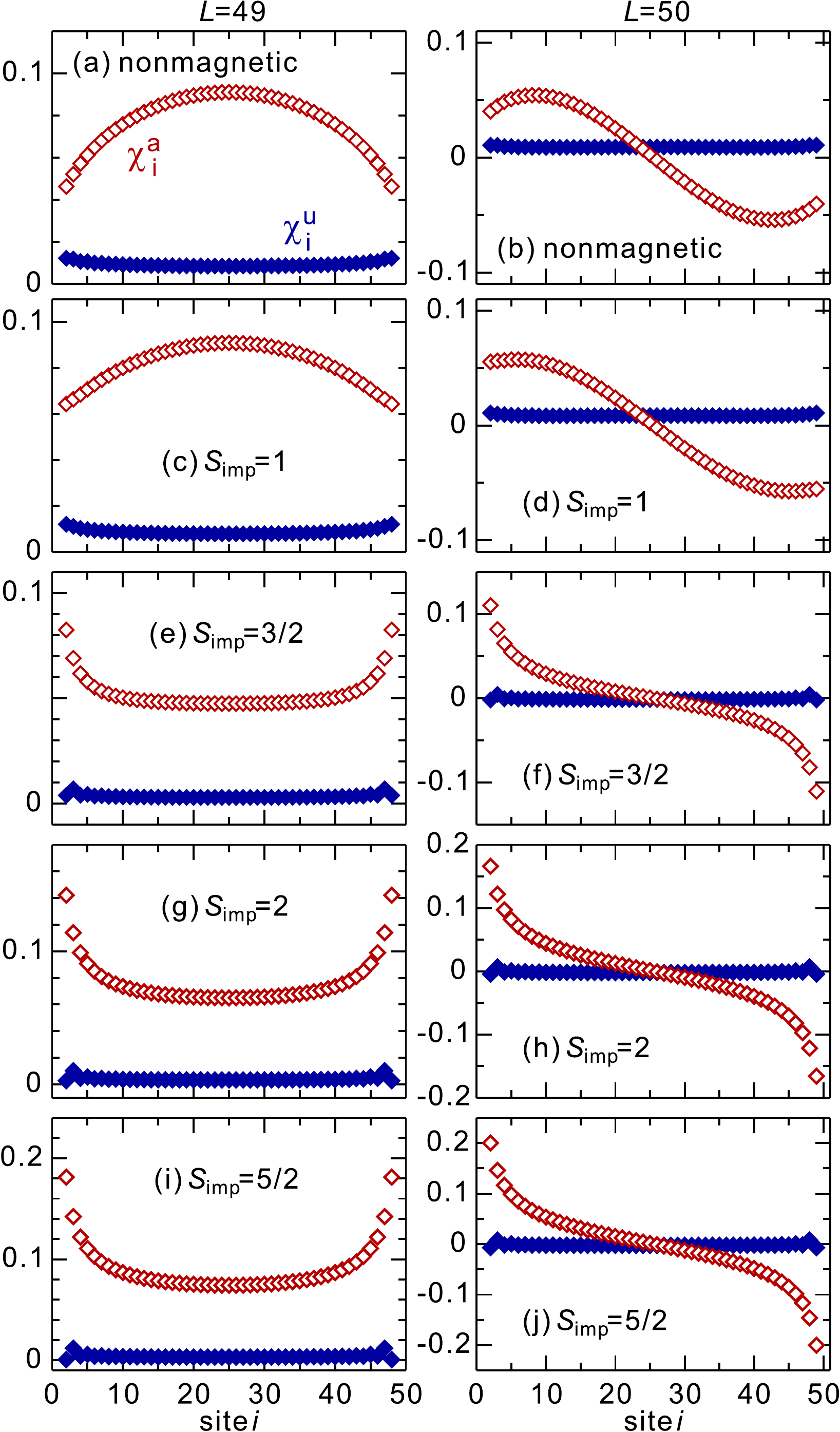}
\caption{
Uniform and staggered components of the local susceptibility $\chi_i$ for (a) $S_{\rm imp}=0$, (b) $S_{\rm imp}=1$, (c) $S_{\rm imp}=3/2$, (d) $S_{\rm imp}=2$, (e) $S_{\rm imp}=5/2$, and (f) $S_{\rm imp}=3$, obtained with spin-$\frac{1}{2}$ chain segments with $49$ (left panel) and $50$ sites (right panel).
}
\label{fig:localchi}
\end{figure}

By studying the local spin susceptibility we can give a theoretical prediction
for measured NMR spectra~\cite{Eggert95,Takigawa97}. The local spin
susceptibility is defined as
\begin{align}
	\chi_i(T)=\frac{1}{T}\sum_{j}\langle S^z_i S^z_j \rangle,
	\label{local_chi}
\end{align}
where $T$ is temperature. Since both the numerator and denominator vanish
in the limit of $T \to 0$, we need to transform this, so that we may calculate $\chi_i(T=0)$.
Following Ref.~\onlinecite{Laukamp98}, the local spin susceptibility at low temperature
is redefined as
\begin{align}
	\chi_i \approx \sum_{j}\langle \phi_S |S^z_i S^z_j| \phi_S \rangle=S\langle \phi_S |S^z_i| \phi_S \rangle,
	\label{local_chi}
\end{align}
where $| \phi_S \rangle$ is the first excited state in the energy spectrum with
$z$-component of total spin $S>0$. Typically, $S$ is 1 and 1/2 for a system
with even and odd number of sites, respectively. The local spin susceptibility is
a symmetric function under reflections with respect to the center of open system
sandwiched by impurities. Since we find only staggered oscillations of $S^z_i$
in the all cases, $\chi_i$ can be divided into a ``uniform'' component $\chi_i^{\rm u}$
and ``staggered'' component $\chi_i^{\rm a}$ using the definition
$\chi_i=\chi_i^{\rm u}-(-1)^i \chi_i^{\rm a}$. The uniform and staggered components
are practically obtained as $\chi_i^{\rm u}=\chi_i/2+(\chi_{i-1}+\chi_{i+1})/4$ and
$\chi_i^{\rm a}=-(-1)^i(\chi_i-\chi_i^{\rm u})$, respectively.

In Fig.~\ref{fig:localchi}(a,b), DMRG results for the uniform and staggered
components of local spin susceptibility using simple open chains with
$49$ and $50$ sites are shown. The open chains are regarded as chain
segments created by doping of nonmagnetic impuritues into an SU(2)
Heisenberg chain. The staggered component $\chi_i^{\rm a}$ is force
to be symmetric for odd $L$ and antisymmetric for even $L$ as a consequence
of finite size effect. As consistent with Ref,~\onlinecite{Laukamp98}, 
the maximum position of $\chi_i^{\rm a}$ is distant as much as possible
from the chain ends. This is also consistent with the prediction that
$\chi_i^{\rm a}$ simply increases as a function of distance from the
nonmagnetic impurity for a semi-infinite chain~\cite{Eggert95}.

Let us then consider the case of magnetic impurity. We first focus on the cases of
$\Delta_{\rm imp}=1$. For this, we need to prepare SU(2) chain segment with
impurities at both ends. We here create such a chain segment by replacing two
opposite sites on an SU(2) periodic chain with two magnetic impurities. Specifically,
periodic SU(2) chains with $100$ and $102$ sites are used to configure chain
segments with $49$ and $50$ sites, respectively. Fig.~\ref{fig:localchi}(c,d) show
DMRG results for $\chi_i^{\rm u}$ and $\chi_i^{\rm a}$ in the case of $S_{\rm imp}=1$
impurity. As might be expected from the above discussions, they are even
quantitatively similar to those in the case of nonmagnetic impurity. This is another
evidence to prove that an $S_{\rm imp}=1$ impurity behaves like a nonmagnetic
impurity. On the other hand, the local spin susceptibility for $S_{\rm imp}>1$
impurities exhibits substantially different features as shown in Fig.~\ref{fig:localchi}(e-j).
The results for $\chi_i^{\rm u}$ and $\chi_i^{\rm a}$ in the cases of
$S_{\rm imp}=3/2$, $2$, and $5/2$ impurities look qualitatively similar.
For both system sizes, $\chi_i^{\rm a}$ has a maximum value near the magnetic
impurities. To illustrate the spin state around the magnetic impurity, as an example,
let us consider the three-spin Heisenberg problem; an $S=3/2$ with two adjacent
$S=1/2$'s. The dominant configuration of the ground state in the $S^z=1/2$ sector
is expressed as
\begin{align}
	\psi\approx|\frac{1}{2},-\frac{1}{2}\rangle \otimes |\frac{3}{2},\frac{3}{2}\rangle \otimes |\frac{1}{2},-\frac{1}{2}\rangle
	\label{phi_loc}
\end{align}
Since the $S=1/2$ states tend to localize at the $S_{\rm imp}=3/2$ impurity,
it leads to the maximum value of $|\chi_i^{\rm a}|$ near the $S_{\rm imp}=3/2$
impurity. Nevertheless, $\chi_i^{\rm a}$ for $L=49$ reduces to a finite saturated
value with the distance from the $S_{\rm imp}=3/2$ impurity. This means that
the localization of $S=1/2$ states near the $S_{\rm imp}>1$ impurities
is not very strong. This trends are also seen in the case of $S_{\rm imp}>0$
impurities. The exact solution of three-spin Heisenberg problem for $S_{\rm imp}>0$
impurities is given in Appendix~\ref{app:threesite}. 

\begin{figure}[tbh]
\centering
\includegraphics[width=1.0\linewidth]{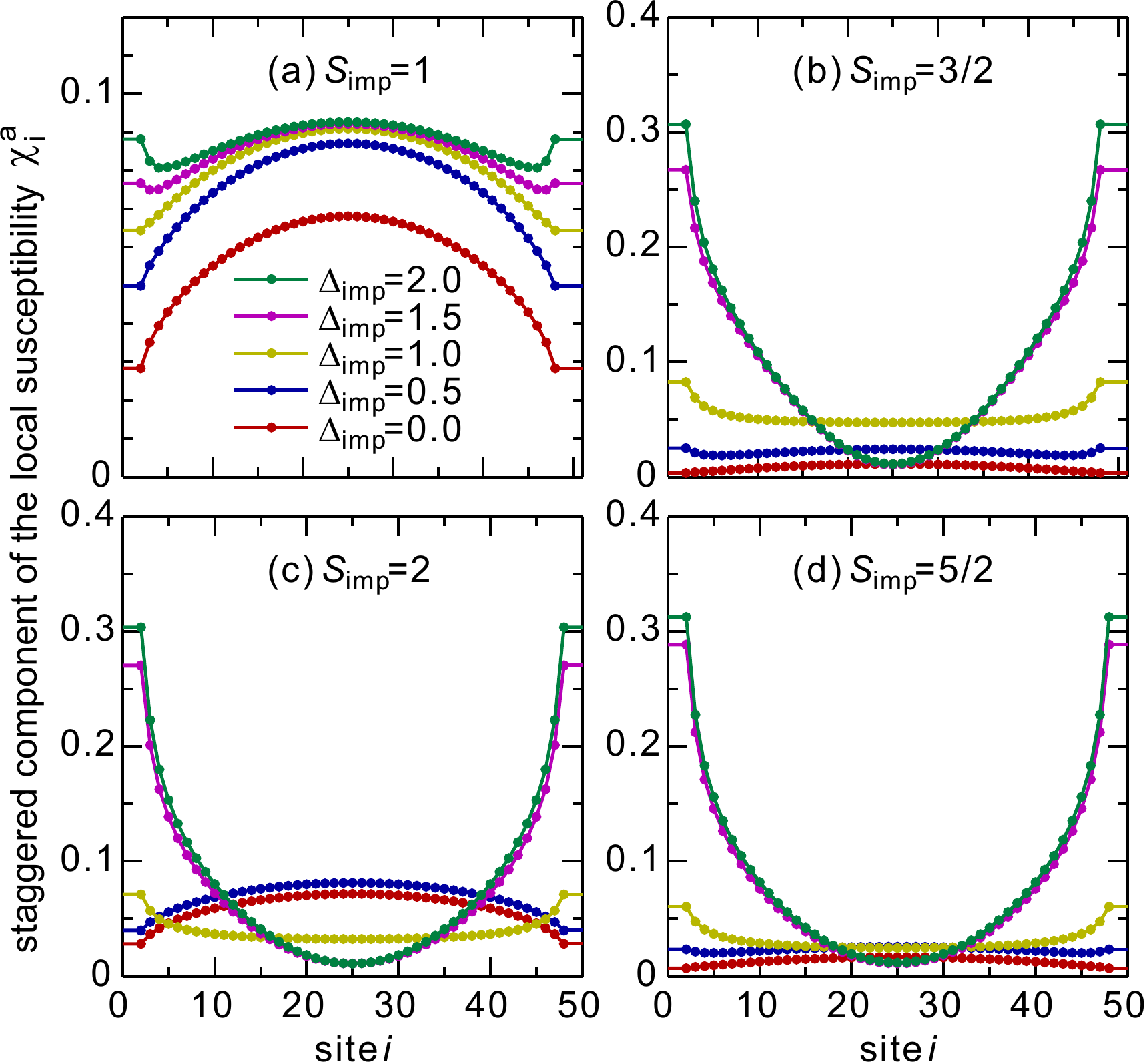}
\caption{
$\Delta_{\rm imp}$-dependence of staggered component of the local
susceptibility $\chi_i$ for (a) $S_{\rm imp}=1$, (b) $S_{\rm imp}=3/2$,
(c) $S_{\rm imp}=2$, and (d) $S_{\rm imp}=5/2$ impurities. The system
length is $L=49$.
}
\label{fig:localchi_delta}
\end{figure}

We further investigate the $\Delta_{\rm imp}$-dependence of $\chi_i^{\rm a}$.
Figure~\ref{fig:localchi_delta}(a) shows DMRG results for $\chi_i^{\rm a}$ at
several $\Delta_{\rm imp}$ values in the case of $S_{\rm imp}=1$ impurity.
The overall trends almost remain unchanged with $\Delta_{\rm imp}$
although $\chi_i^{\rm a}$ is broadly enhanced (suppressed) with increasing
(decreasing) $\Delta_{\rm imp}$. This may be naively expected from the fact
that $S_{\rm imp}=1$ impurity behaves like nonmagnetic impurity for any
$\Delta_{\rm imp}$. On the other hand, as shown in Fig.~\ref{fig:localchi_delta}(b-d),
$\chi_i^{\rm a}$ for $S_{\rm imp}>1$ impurities is significantly affected by
$\Delta_{\rm imp}$. With increasing $\Delta_{\rm imp}$, $\chi_i^{\rm a}$
near the magnetic impurity is markedly enhanced and it becomes
very small at the middle of chain. This is interpreted to mean that the
$S=1/2$ states are strongly localized near the $S_{\rm imp}>1$ impurity
because the state denoted by Eq.~\eqref{phi_loc} becomes progressively
more dominant for larger $\Delta_{\rm imp}$. While at small $\Delta_{\rm imp}<1$
the localization of $S=1/2$ states near the magnetic impurity is weakened
(also see Appendix~\ref{app:threesite}), $\chi_i^{\rm a}$ has its maximum
at the middle of chain like in the cases of nonmagnetic and $S_{\rm imp}=1$
impurities. This effect is most pronounced in the case of $S_{\rm imp}=2$.
Incidentally, in the case of $S_{\rm imp}=1/2$ impurity, the local spin
susceptibility is always very small for $0<\Delta_{\rm imp}<5$ and thus
$|\chi_i^{\rm a}| \lesssim 0.01$.

\begin{figure}[tbh]
\centering
\includegraphics[width=1.0\linewidth]{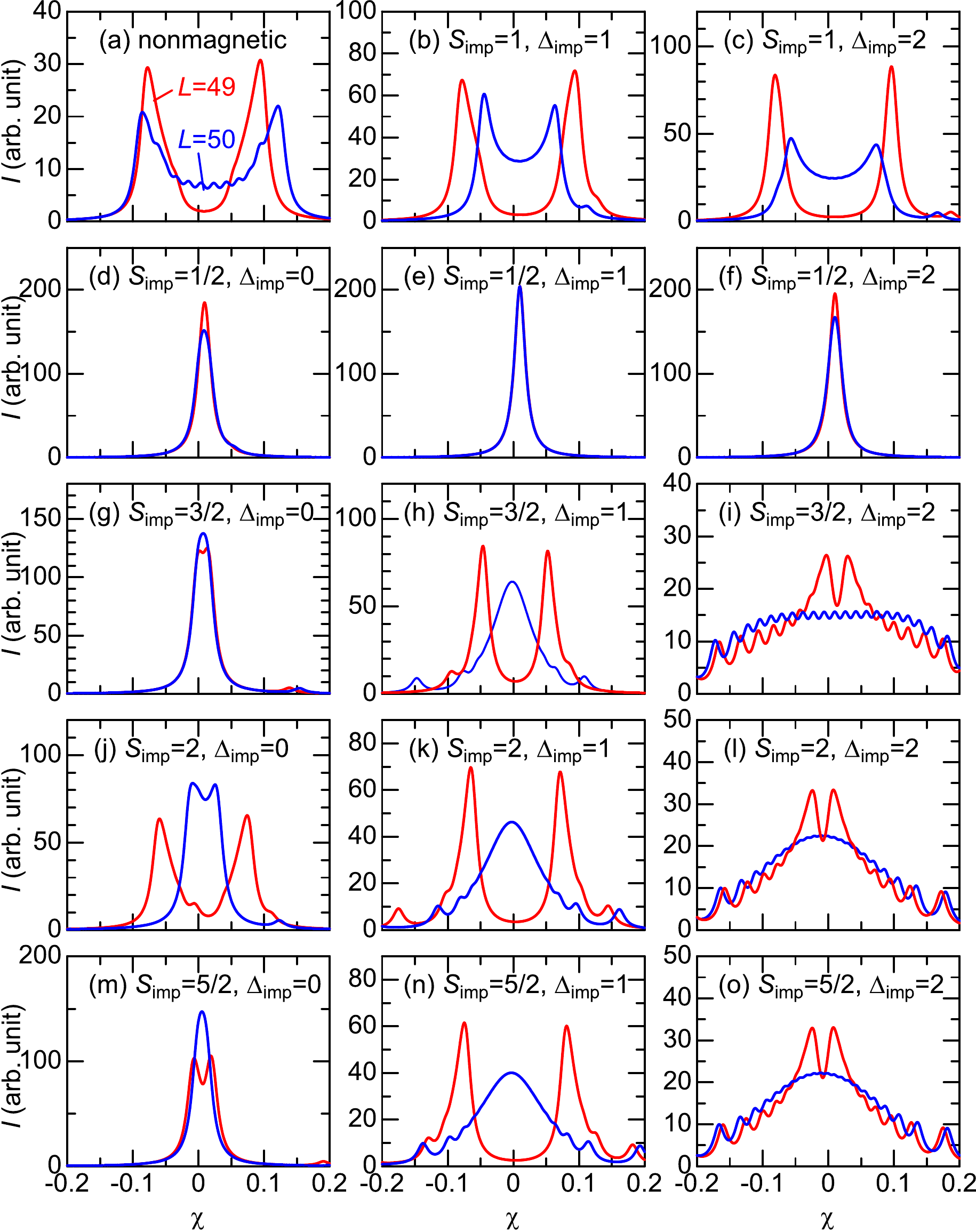}
\caption{
$\Delta_{\rm imp}$-dependence of NMR spectra for (a) $S_{\rm imp}=0$,
(b,c) $S_{\rm imp}=1$, (d-f) $S_{\rm imp}=1/2$, (g-i) $S_{\rm imp}=3/2$.
(j-l) $S_{\rm imp}=2$. and (m-o) $S_{\rm imp}=5/2$ impurities.
A Lorentzian broadening $0.01$ is introduced.
}
\label{fig:nmr}
\end{figure}

Once the staggered component of the local susceptibility is calculated,
it would be interesting to see its amplitude distribution, which can be
detected as a broadening of the NMR spectra~\cite{Eggert95,Takigawa97}.
A Lorentzian broadening $0.01$ is introduced to obtain the NMR spectra.
Figure~\ref{fig:nmr}(a-c) shows the NMR spectra for nonmagnetic and
$S_{\rm imp}=1$ impurities. In these cases, we find a two-peak structure,
where the distance between peaks roughly corresponds to an amplitude
of $\chi_i$, because the maximum position of $\chi_i^{\rm a}$ moves away
from the impurities due to the delocalized character of spinon excitation. 
For the $S_{\rm imp}=1/2$ impurity, as shown in Fig.~\ref{fig:nmr}(d-f),
the NMR spectra always consist of a single sharp peak because of 
$|\chi_i^{\rm a}| \lesssim 0.01$ for any $\Delta_{\rm imp}$. 

Based on the results for $\chi_i^{\rm a}$, a strong dependence of
NMR spectra on $\Delta_{\rm imp}$ is expected for $S_{\rm imp}>1$
impurities. The NMR spectra for $S_{\rm imp}=3/2$, $2$, and $5/2$
impurities are shown in Fig.~\ref{fig:nmr}(g-o). The results in these
three cases are qualitatively similar: At $\Delta_{\rm imp}=1$
the NMR spectra for $L=50$ exhibits a broad peak centered at zero,
reflecting a weak localization of the $S=1/2$ states near the magnetic
impurities. At $\Delta_{\rm imp}=2$ the localization of $S=1/2$ states
near the magnetic impurities is strongly enhanced and the NMR spectra
becomes much broader. At $\Delta_{\rm imp}=0$ the NMR spectra
show a rather narrow peak because the amplitude of $\chi_i^{\rm a}$ is
very small like in the case of $S_{\rm imp}=1/2$ impurity. 

We then analyze the experimental NMR spectra of the Co-doped
SrCuO$_2$ compound SrCu$_{0.99}$Co$_{0.01}$O$_2$~\cite{Utz2017}. 
A very broad peak with a central small dip was experimentally
observed at low temperature. A remaining issue is that the spin state
of Co ion, which is either $S=3/2$ or $S=1/2$, has not been identified
up to now. We could find an answer to this question, although the results
for $L=49$ and $L=50$ may reveal typical features of $\sim 2\%$ doping
and we need to take an average over $L$ using Eq.~\eqref{probability}
with $n_{\rm imp}=0.01$ for quantitative analysis. We argue that
the low-temperature broad NMR spectra SrCu$_{0.99}$Co$_{0.01}$O$_2$
can be explained only when the spin state of Co ion is assumed to be
$S=3/2$ and the spin anisotropy to be $\Delta_{\rm imp}>1$.

\subsection{Staggered magnetization with finite impurity doping}\label{sec:mag}

\begin{figure}[tb]
\centering
\includegraphics[width=0.9\linewidth]{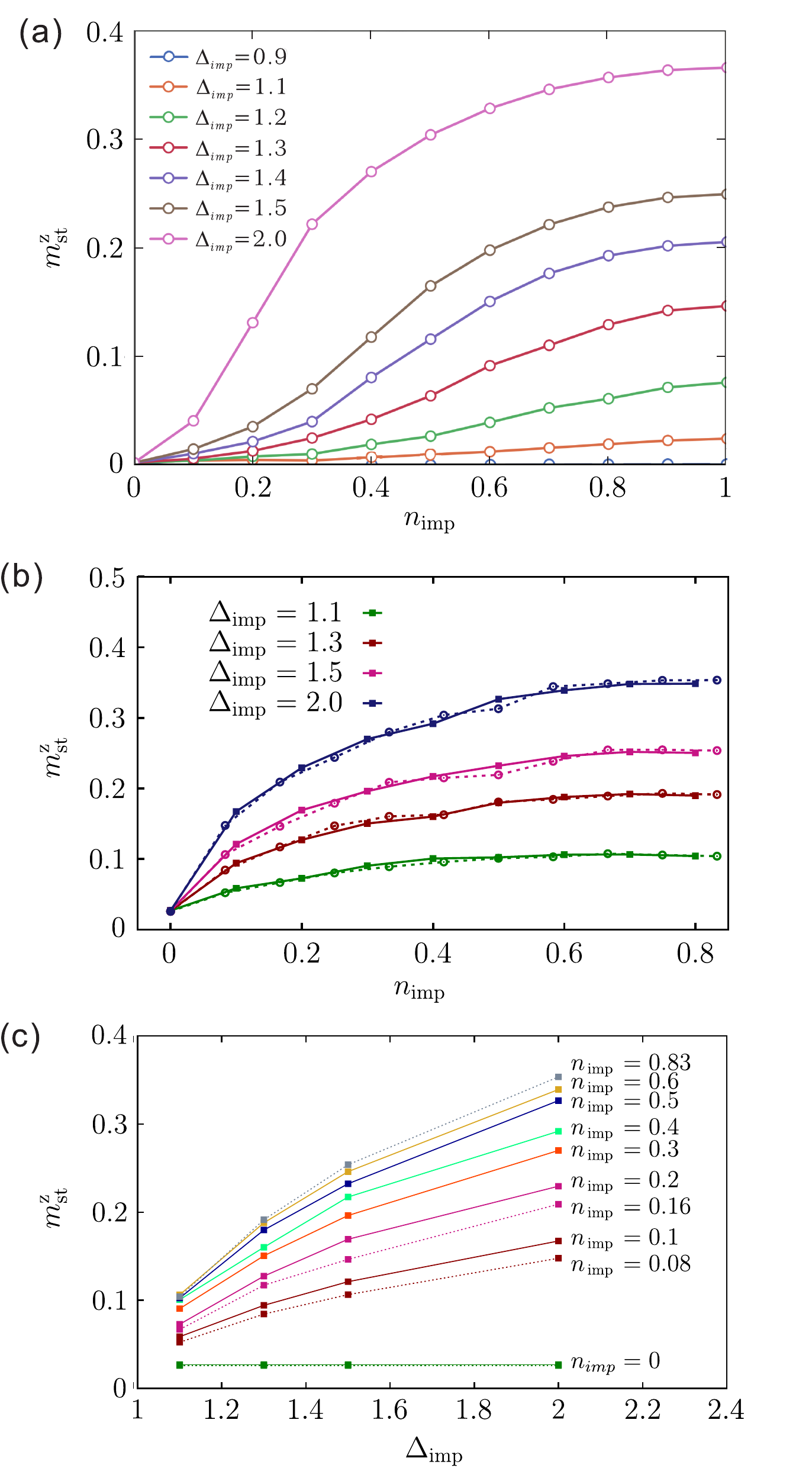}
\caption{
Variation of staggered magnetization $m_z^{\rm st}$ with impurity density
$n_{\rm imp}$ for different values of anisotropy ($\Delta_{\rm imp}$) with using
(a) DMRG and (b) CMFT approaches. (c) CMFT results for $m_z^{\rm st}$ as
a function of $\Delta_{\rm imp}$ at fixed values of $n_{\rm imp}$. In (b ) and (c),
the solid and dashed lines correspond to the CMFT results for $N_c=10$ and $12$,
respectively.
}
\label{fig:stag_mz}
\end{figure}

Let us then investigate the staggered magnetization $m_{\rm st}^z$, defined by
Eq.~\eqref{eq:stm}, as functions of the XXZ anisotropy $\Delta_{\rm imp}$ and
impurity density $n_{\rm imp}$. As mentioned above, the effect of magnetic
impurity on the stability of N\'eel order in the SU(2) Heisenberg chain can be
classified broadly into two types: One is nonmagnetic and $S_{\rm imp}=1$
impurities which simply suppress the N\'eel stability and the other
is $S_{\rm imp}=1/2$ and $S_{\rm imp}>1$ impurities which encourage
the development of long-range N\'eel order if $\Delta_{\rm imp}\gtrsim1$.
Thus, it would be reasonable to consider the case of $S_{\rm imp}=1/2$
as a representative of the latter although there may be quantitative differences
from the case of $S_{\rm imp}>1$. Therefore, we focus on the case of
$S_{\rm imp}=1/2$ impurity hereafter.

When an SU(2) Heisenberg chain is doped with an $S_{\rm imp}=1/2$
impurity with $\Delta_{\rm imp}>1$, a staggered oscillation
of $\langle S^z_i \rangle$ may be induced as a Friedel oscillation from
the impurity. The amplitude of this Friedel oscillation decays as
$\langle S^z_r \rangle \sim 1/r^\eta$ since it mimics the decay of
spin-spin correlation function of the SU(2) Heisenberg chain~\cite{White2002}.
The value of $\eta$ can be estimated to be $\sim 1$ from the slope of
$C_{\rm s}(1,r)$ in Fig.~\ref{fig:szsz_delta}(a). This means that, if the SU(2)
chain is doped with more than one impurity, the staggered oscillation reaches
the next impurity with maintaining its amplitude of the order of
$C_s(1,1)/\bar{l} \approx 0.2n_{\rm imp}/(1-n_{\rm imp})$. As shown in
Fig.~\ref{fig:dejtadepS12}(d), this staggered oscillation is further enhanced
at the next impurity. Therefore, $m_{\rm st}^z>0$ is naively expected as long
as $n_{\rm imp}>0$ and $\Delta_{\rm imp}>1$. In other words, the presence
of finite amount of XXZ impurities with $\Delta_{\rm imp}>1$ induces
a long-range N\'eel order in the doped SU(2) Heisenberg chain.
In order to numerically confirm it, we calculate $m_{\rm st}^z$ for various
sets of $\Delta_{\rm imp}$ and $n_{\rm imp}$ using DMRG method.
We study open chains with length $N_c=40-800$. For a given length $N_c$,
the value of $m_{\rm st}^z$ is obtained by averaging over $10000/N_c$
samples, and then a finite-size scaling analysis to the thermodynamic limit
$N_c \to \infty$ is performed. In Fig.~\ref{fig:stag_mz}(a), the obtained
values of $m_{\rm st}^z$ are plotted as a function of $n_{\rm imp}$
for several values of $\Delta_{\rm imp}$. We find that $m_{\rm st}^z$
is always finite for $n_{\rm imp}>0$	 and $\Delta_{\rm imp}>1$.
The qualitative trend of $m_{\rm st}^z$ vs. $n_{\rm imp}$ is common for any
$\Delta_{\rm imp}(>1)$; it is small at low $n_{\rm imp}$, increases rapidly at some
intermediate $n_{\rm imp}$, and saturates to a value at high $n_{\rm imp}$.
The saturation value agrees perfectly with spontaneous magnetization of the
XXZ Heisenberg chain~\cite{Yang1966-1,Yang1966-2}. 
For a fixed $n_{\rm imp}(>0)$, $m_{\rm st}^z$ increases with increasing
$\Delta_{\rm imp}$. This is a natural consequence of the fact that the
amplitude of induced staggered oscillation is larger for larger $\Delta_{\rm imp}$.
It is interesting that the largest slope of $\partial m_{\rm st}^z/\partial n_{\rm imp}$
is given by smaller $n_{\rm imp}$ for larger $\Delta_{\rm imp}$.

We also estimate $m_{\rm st}^z$ using CMFT, which may provide a cross-check
for the above DMRG results. The ground state of an AFM spin chain as obtained
by CMFT is a mixed state forming valence bonds on alternate sites in the presence
of very small yet finite (less than $10\%$ of the spin magnitude) staggered
magnetization. The presence of an impurity introduces $\Delta_{\rm imp}$
in the bonds connecting the impurity. In Fig~\ref{fig:stag_mz}(b), we show the CMFT
results for variation of $m_{\rm st}^z$ with $n_{\rm imp}$ for $N_c=10$ (solid lines),
and $12$ (dashed lines) averaged over $20$ random configurations. For a particular
$\Delta_{\rm imp}$, increase in the number of anisotropic bonds increases $m_{\rm st}^z$,
indicating the formation of stronger N\'eel order. It is interesting to note that anisotropic
strength as low as $n_{\rm imp}=0.1$ can induce an ordered state.
Since calculations performed on larger cluster size of $N_c=12$ (dashed lines)
do not show any qualitative difference in the results, most of the analysis will
be restricted to a cluster of 10 sites.

Figure~\ref{fig:stag_mz}(c) shows the evolution of $m_{\rm st}^z$ for different
$n_{\rm imp}$ with increasing $\Delta_{\rm imp}$. It suggests that introducing
a single impurity spin ($n_{\rm imp}=0.1$ in a spin chain of $N_c=10$ and
$n_{\rm imp}=0.08$  in a spin chain of $N_c=12$) affects the overall order
of the chain. This effect can be further increased by increasing
$\Delta_{\rm imp}$. Note that configurations do not include edge spins so
MF bonds are always pure. It is interesting to observe that staggered
magnetization increases (decreases) with increasing (decreasing) the cluster size
since this sampling underestimates (overestimates) magnetization for lower
(higher) impurity density. Detailed discussion on the number of
random samplings is given in Appendix~\ref{app:randomsampling}.
While the qualitative behavior of $m_{\rm st}^z$ calculated using CMFT matches
very well with that obtained via DMRG, results obtained using CMFT suffer
from finite-size mean-field effects. The finite-size effect leads to higher $m_{\rm st}^z$
in comparison to DMRG. This effect is more pronounced for small $n_{\rm imp}$,
as for low $n_{\rm imp}$ the number of Heisenberg bonds is larger and CMFT
breaks the rotational symmetry of the Hamiltonian. CMFT results can be further
improved by increasing the cluster size. Detailed discussion on the finite-size
effect is given in Appendix~\ref{app:sizeeffect}.

\subsection{Uniform magnetization with external field}\label{sec:mag_hz}

\begin{figure}[t]
\centering
\includegraphics[width=1.0\linewidth]{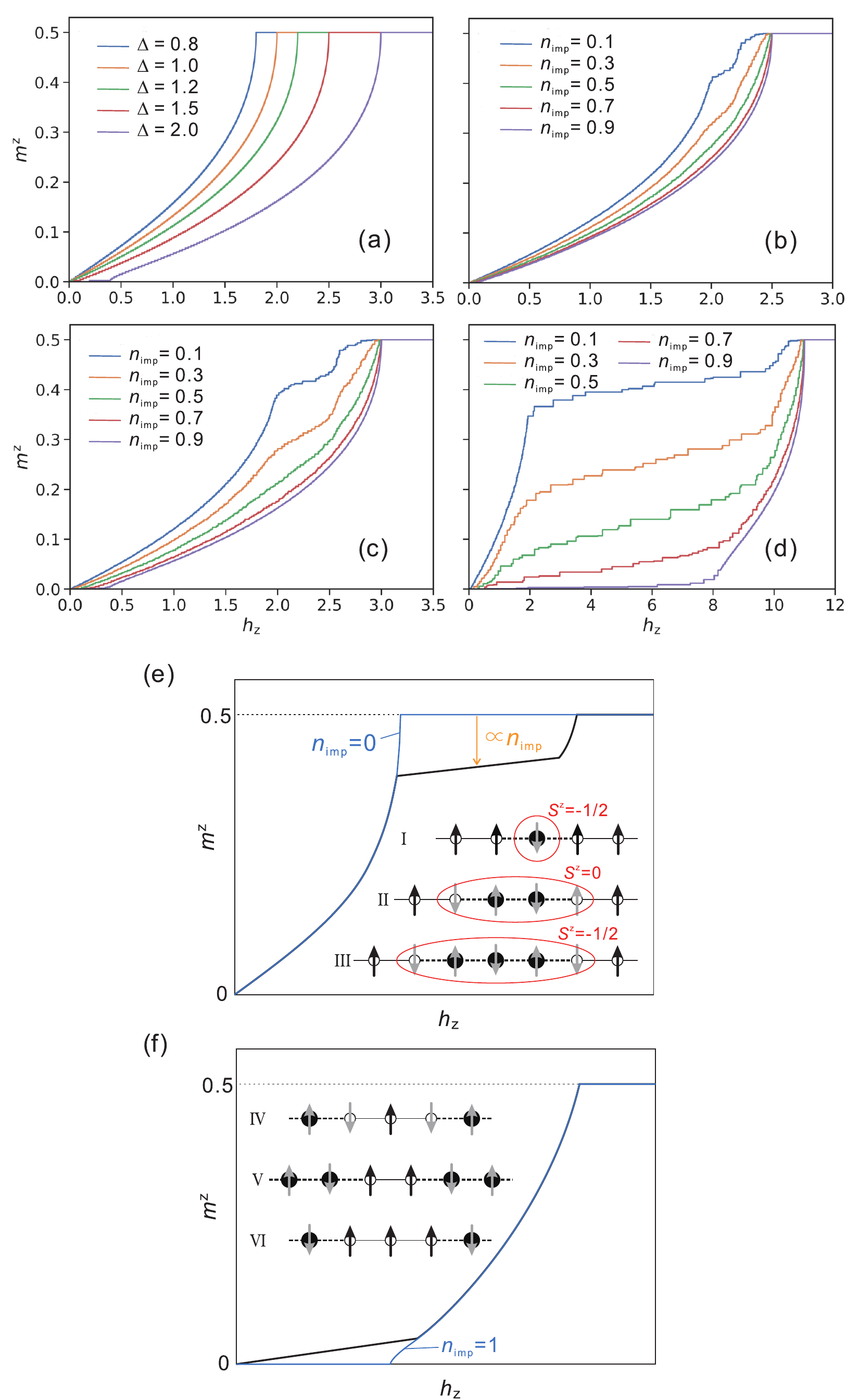}
\caption{
(a) Magnetization curve of XXZ Heisenberg chain as a function of magnetic
field $h_z$. (b-d) Magnetization curves of doped SU(2) Heisenberg chain
by $S_{\rm imp}=1/2$ XXZ impurities with (b) $\Delta_{\rm imp}=1.5$,
(c) $2$, and (d) $10$ for several impurity densities $n_{\rm imp}$.
(e,f) Schematic illustrations of magnetization curve near (e) $n_{\rm imp}=0$
and (f) $n_{\rm imp}=1$. The insets of (e) [(f)] show schematic representations
of typical spin states near $n_{\rm imp}=0$ at high fields [$n_{\rm imp}=1$ at
low fields], where the solid line denotes spin-isotropic XXX bond and the dashed
line denotes spin-anisotropic XXZ bond.
}
\label{fig:mag_dmrg}
\end{figure}

In this subsection, we discuss the $\Delta_{\rm imp}$ and $n_{\rm imp}$
dependence of total magnetization in the presence of external field $h_z$.
The total magnetization is defined by
\begin{align}
m^z=\frac{1}{L}\sum_1^L \langle S^z_i \rangle.
	\label{eq:totm}
\end{align}
We begin by discussing the behavior of uniform magnetization $m^z$ of
pure XXZ Heisenberg chain with exchange anisotropy $\Delta$ as
a function of $h_z$. The Hamiltonian reads
${\cal H} =\sum_i [(S^+_iS^-_{i+1}+S^-_iS^+_{i+1})/2+\Delta S^z_iS^z_{i+1}]+h_z\sum_i S^z_i$.
The DMRG results for $m^z$ are plotted in Fig.~\ref{fig:mag_dmrg} (a).
For $\Delta \le 1$, with increasing $h_z$, $m^z$ increases almost linearly at low $h_z$ and
exhibits a divergent saturation reflecting the strong quantum fluctuations.
For $\Delta > 1$, $m^z$ remains zero up to a finite $h_z(\equiv h_{z,{\rm cr}})$
because of the long-range N\'eel order, it rises up vertically at $h_z=h_{z,{\rm cr}}$,
and at higher $h_z$ behavior is qualitatively similar to that for $\Delta \le 1$.
The saturation field is $h_{\rm sat}=1+\Delta$ for any $\Delta$.
We then investigate the $n_{\rm imp}$ dependence of magnetization curve
for doped SU(2) Heisenberg chain by XXZ anisotropic $S_{\rm imp}=1/2$
impurities with $\Delta_{\rm imp}$.
In other words, we see how the magnetization curve is varied from that
of $\Delta=1$ to that of $\Delta=\Delta_{\rm imp}$ with increasing
$n_{\rm imp}$.

In Fig.~\ref{fig:mag_dmrg} (b-d), DMRG results for the magnetization
curve with $\Delta_{\rm imp}=1.5$, $2$, and $10$ at several values of
$n_{\rm imp}$ are shown. The undoped system is an SU(2) Heisenberg chain.
For each curve shown in Fig.~\ref{fig:mag_dmrg} (b-d), $m^z$ is
calculated by taking average over 10 random realizations of impurity distributions
on an open chain of length $N_c=1000$. Qualitatively, the $n_{\rm imp}$
dependence of $m^z$ vs $h_z$ is similar for all $\Delta_{\rm imp}$ cases.
Nevertheless, the characteristics become more obvious with larger $\Delta_{\rm imp}$.
The specific features can be interpreted by the fact that XXX spins are more easily
polarized than XXZ spins with easy-axis anisotropy ($\Delta_{\rm imp}>1$).

At low impurity density ($n_{\rm imp}\sim0$), a plateau-like feature is seen near
the saturation $m^z\sim1/2$. If the impurities are dilute, the system consists
of field-polarized spins and isolated magnons at high fields because the XXZ
impurity spins are initially flipped with decreasing field from the saturation.
This spin state is sketched as I in the inset of Fig.~\ref{fig:mag_dmrg}(e).
Since there is a `gap' between field strengths to polarize the XXZ impurity spins 
and the other spins, a plateau is created at $m^z \approx (1-n_{\rm imp})/2$.
As $n_{\rm imp}$ increases, the probability that two or more XXZ impurities
are aligned increases in the $h_z$ range of plateau. For example, when
two XXZ impurities are placed next to each other, as II in the inset of
Fig.~\ref{fig:mag_dmrg}(e), a tetramer singlet ($S^z=0$) is composed of four
spins connected by XXZ coupling (we call a chain segment like the
tetramer in this case ``XXZ chain segment''). Thus, a finite gap to $S^z=1$
takes the shape of plateau. Similarly, when three XXZ impurities are aligned,
a pentamer is formed as III in the inset of Fig.~\ref{fig:mag_dmrg}(e).
Having polarized spins on either side of the pentamer, a gap opens between
$S^z=-1/2$ and $S^z=1/2$ states. It also creates a plateau. In general,
an XXZ chain segment with even (odd) number of XXZ impurities exhibits
a gap between $S^z=0$ and $S^z=1$ states ($S^z=-1/2$ and $S^z=1/2$ states).
Since the gap value differs by the length of XXZ chain segment, the plateau
has a finite slope. Basically, odd-length XXZ chain segments have larger gap
than even-length ones; the gap is larger for shorter XXZ chain segment.
The more various lengths of XXZ chain segments exist, the more the slope of
plateau becomes steep. Indeed, the increase of slope with increasing $n_{\rm imp}$
can be confirmed up to $n_{\rm imp}=0.3$ in Fig.~\ref{fig:mag_dmrg}(b-d).

At high impurity density ($n_{\rm imp}\sim1$) the overall shape of $m^z$ vs.
$h_z$ looks similar to that of XXZ Heisenberg chains with $\Delta=\Delta_{\rm imp}$
because most parts of the system are covered by XXZ chain segments
with $\Delta_{\rm imp}$. However, a slow increase of $m^z$ is seen
at lower $h_z$ instead of gapped behavior in the XXZ Heisenberg chain.
This is caused by a scattering of undoped XXX chain segments, as sketched
in the inset of Fig.~\ref{fig:mag_dmrg}(f). The XXX chain segments are more
easily polarized than XXZ chain segments. Since typical field strengths
exhibiting the states IV, V, and VI are different, i.e.,
$h_z({\rm VI})>h_z({\rm V})>h_z({\rm IV})$, a gradual slope of $m^z$
at lower $h_z$ is created.

\begin{figure}[tbh]
\centering
\includegraphics[width=1.0\linewidth]{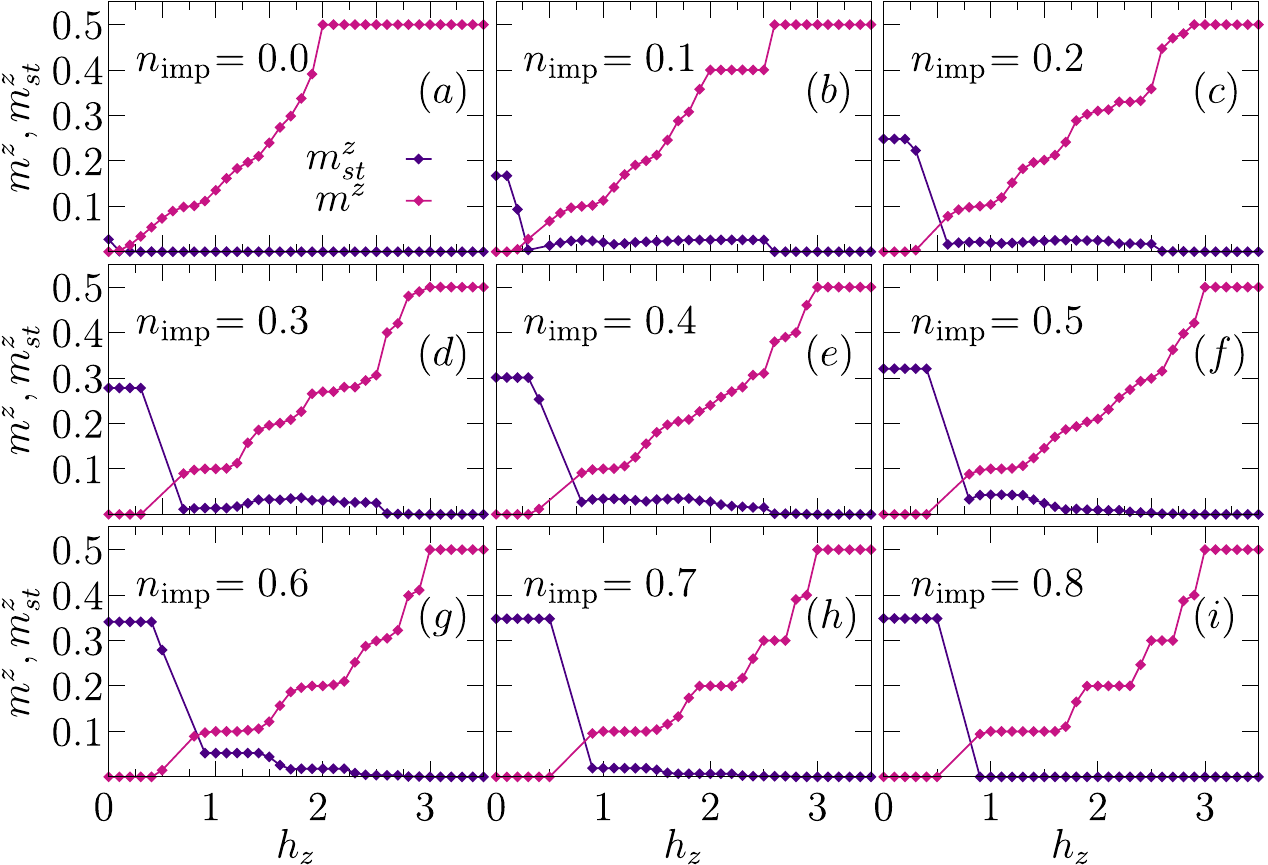}
\caption{
Variations of total magnetization $m^z$ (pink) and staggered magnetization
$m_{\rm st}^z$ (purple) with increasing applied magnetic field for different
values of $n_{\rm imp}$ as indicated in the panels. $\Delta$ is fixed to 2 and
$N_{av}$ is set to 10.
}
\label{fig:mag_CMFT}
\end{figure}

It would be informative to compare CMFT results for the uniform magnetization
$m^z$ with those by DMRG. In Fig.~\ref{fig:mag_CMFT}, we show CMFT results
for $m^z$ as a function of applied magnetic field for different values of impurity densities.
Similar to the DMRG results, for $n_{\rm imp}=0$, total magnetization continuously
increases with applied field until saturation where all the spins align in the direction
of field [see Fig.~\ref{fig:mag_CMFT} (a)]. Small step-like features obtained in net
magnetization is a consequence of finite-size effects in CMFT, which can be
reduced with increasing cluster size. Numerically exact results obtained from
DMRG, begins from zero magnetization owing to gapless nature of 1D Heisenberg
AFM and continuously increases until saturation field. While the intermediate
behavior of $m^z$ obtained by CMFT does not match with the results of DMRG,
it is interesting to note that the value of saturation field agrees very well.

In Fig.~\ref{fig:mag_CMFT} (b-i) we show the variation of total magnetization
$m^z$ (pink) and staggered magnetization $m_{\rm st}^z$ (purple) with increasing
applied magnetic field for different values of $n_{\rm imp}$ as indicated in the panels.
A N\'eel order is formed on substitution of impurities, which vanishes above
a finite magnetic field. This critical value of the field above which N\'eel order
vanishes increases with increase in $n_{\rm imp}$. This is understood as
the closing of the gap existing above the ground state. This feature is reflected
in $m^z$ remaining zero up to a finite strength of magnetic field that is equal
to the energy gap above the ground state. In the XXZ limit, this gap is related
to the anisotropy in Hamiltonian~\cite{Wessel2000,Des_Cloizeaux1966}.
Note that results for some field values are omitted as for those applied field
values, mean fields did not converge for some random configurations.
It is interesting as mean fields converged for all other values of applied field
with the same set of random configurations. Further increase in the impurity
density increases the saturation magnetic field strength. For the extreme
case, when $n_{\rm imp}=0.8$, all the bonds except mean-field decoupled ones
are anisotropic in nature. Disregarding the field values of non-convergence,
the behavior of net magnetization and staggered magnetization reveal formation
of N\'eel order in low fields and fully saturated state in high field limit.
The results for $m^z$ with applied field matches with those obtained in
BaCo$_2$V$_2$O$_8$~\cite{Kimura2008-2}. This material is expected to be
explained by a quasi-1D XXZ model with $\Delta_{\rm imp}=2$. An ideal XXZ
chain model will show a quantum phase transition from the N\'eel ordered
phase to Tomonaga-Luttinger liquid to saturated state at high fields.
Further investigations are required to understand this intermediate order
appearing in the presence of field using CMFT.

\section{Thermodynamic properties}

In previous section, we have established that the antiferromagnetic order is
induced in an isotropic spin chain on substitution of a spin-1/2 magnetic impurity
resulting in anisotropic neighboring bonds. We now discuss signatures
of phase transition appearing in specific heat ($C_v$) and susceptibility ($\chi$)
with increasing temperature for different values of impurity density $n_{\rm imp}$
and $\Delta_{\rm imp}$. 

\begin{figure}[tbh]
\centering
\includegraphics[width=0.9\linewidth]{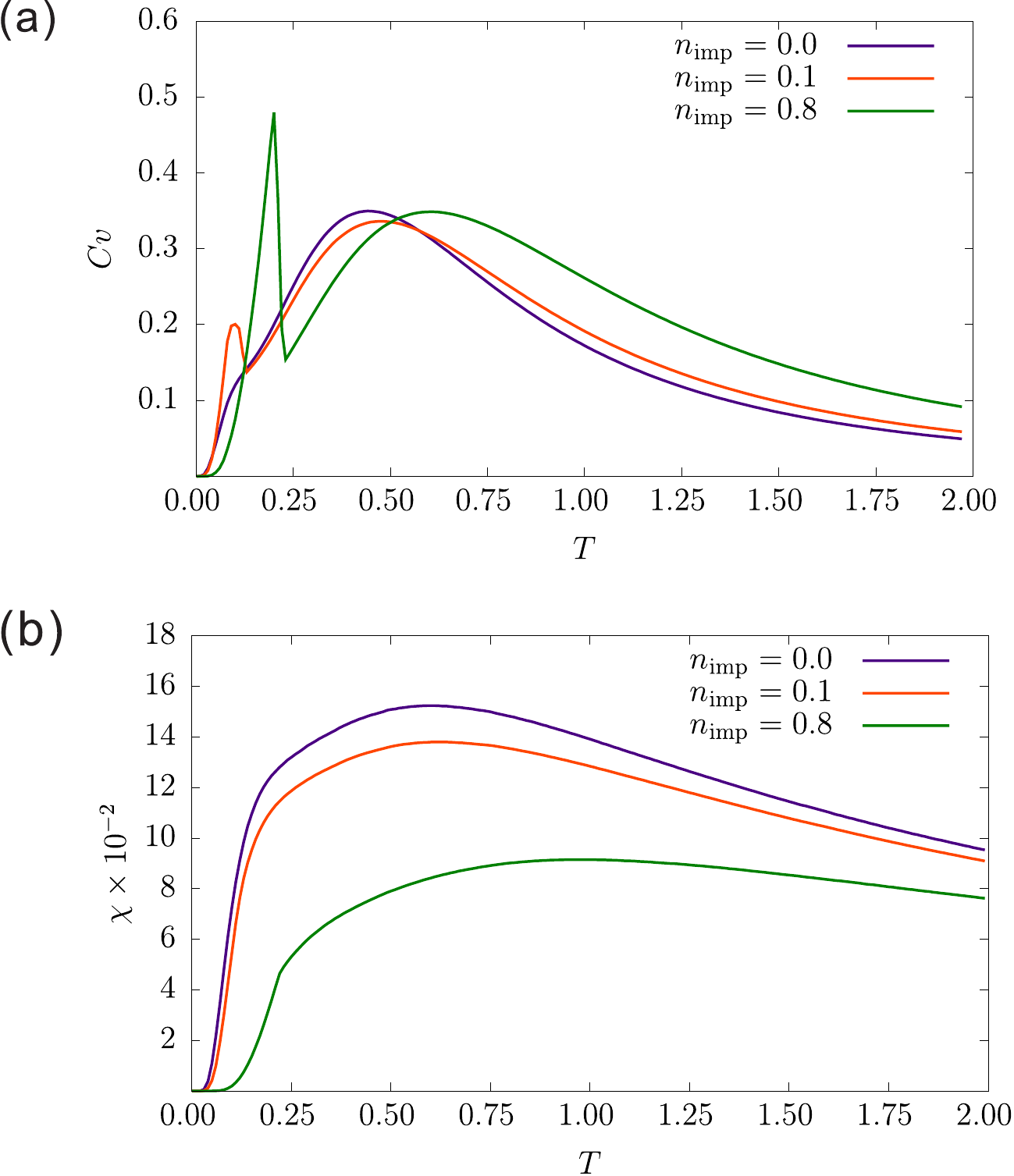}
\caption{ CMFT results for variation of (a) specific heat and (b) susceptibility
with temperature for different values of impurity density while the anisotropy
strength is fixed to $\Delta_{\rm imp}=2$.
}
\label{fig:thermo}
\end{figure}

Figure~\ref{fig:thermo} (a) shows variation of $C_v$ with temperature for different
$n_{\rm imp}$ with $\Delta_{\rm imp}=2$. For a completely isotropic spin chain
($n_{\rm imp}=0$), specific heat changes smoothly with broadened peak near
$T=0.4$. For impurity density as low as $n_{\rm imp}=0.1$, a small peak emerges
as a consequence of phase transition from a N\'eel phase to a paramagnet.
The peak in $C_v$ sharpens when $n_{\rm imp}$ increases. The huge bump
in $C_v$ for higher temperature is a consequence of continuously decreasing
correlations among spins. The origin of this behavior is discussed in detail in
Sec.~\ref{sec:mag_hz}. Figure~\ref{fig:thermo}(b) shows that $\chi$ decreases
with increase in impurity density. Results for susceptibility in the completely
Heisenberg limit show an unusual curvature in $T<0.25$, which
differs from the behavior expected from the ideal 1D antiferromagnetic Heisenberg
chain. The result shown here is a consequence of the mixed N\'eel order
with dimer correlations as obtained using CMFT. The susceptibility matches
with the one expected from an alternating exchange antiferromagnetic chain
(weakly coupled dimers)~\cite{Johnston2000}. A significant deviation appears 
for high $n_{\rm imp}$ in lower temperature region while the behavior qualitatively
remains the same in higher $T$ limit.

\begin{figure}[tbh]
\centering
\includegraphics[width=1.0\linewidth]{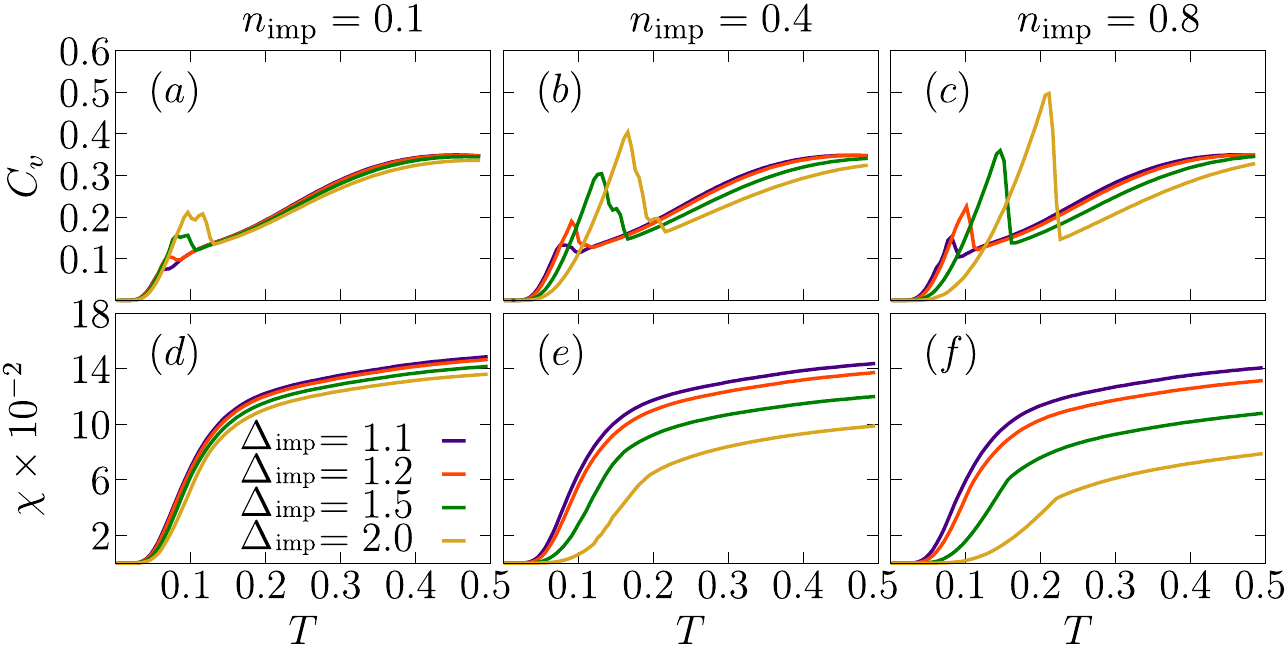}
\caption{
Variations of specific heat and susceptibility for different values of
$\Delta_{\rm imp}$. The results are obtained for (a,d) $n_{\rm imp}=0.1$,
(b,e) $n_{\rm imp}= 0.4$, and (c,f) $n_{\rm imp}= 0.8$.
}
\label{fig:cv_chi}
\end{figure}

In Fig.~\ref{fig:cv_chi}, we show the dependence of $\Delta_{\rm imp}$ on
specific heat and susceptibility for different impurity densities. Results for specific
heat reveal that the increase in the anisotropy strength ($\Delta_{\rm imp}$)
increases the transition temperature as well as the specific heat peak.
This effect is consistent for all impurity densities $n_{\rm imp}=0.1$, $0.4$,
and $0.8$ as shown in Fig.~\ref{fig:cv_chi}(a-c). Results for the susceptibility
in completely isotropic case ($n_{\rm imp}=0$) calculated using CMFT is
similar to that of $\chi$ obtained for antiferromagnetic dimer
chain~\cite{Johnston2000}. For $n_{\rm imp}=0.1$, no significant change in
susceptibility is identified on increasing $\Delta_{\rm imp}$. Further increase
in impurity density leads to decrease in $\chi$, which is prominent for
higher $\Delta_{\rm imp}$.

\begin{figure}[tbh]
\centering
\includegraphics[width=1.0\linewidth]{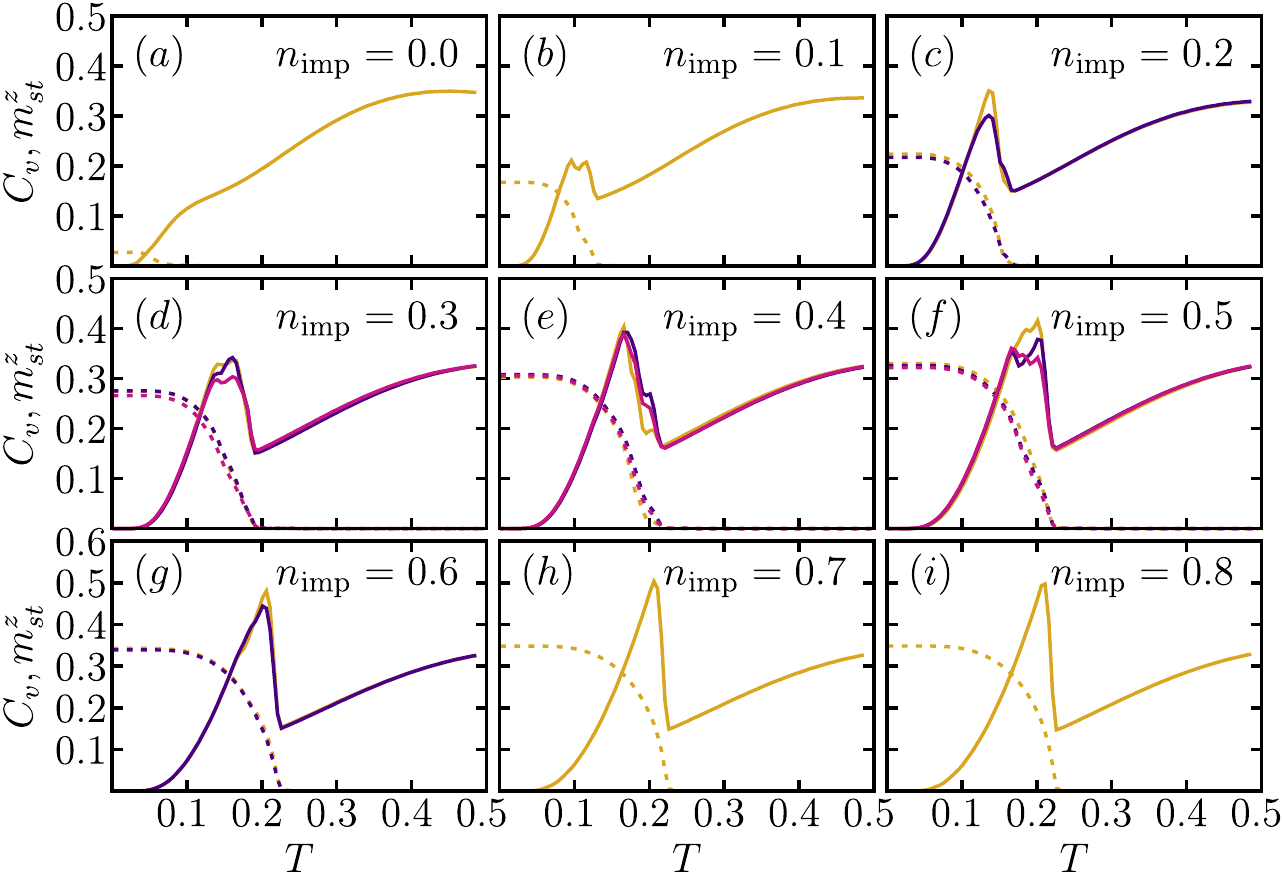}
\caption{
Variations of specific heat (solid lines) and staggered magnetization (dashed lines)
with temperature obtained using CMFT on a cluster size consisting of 10 spins
 Dependence on $n_{imp}$ is shown for various number of random configurations;
 $N_{av}$= 10 (yellow), 20 (blue), and 40 (red).
 }
\label{fig:Cv_n}
\end{figure}

\begin{figure}[tbh]
	\centering
	\includegraphics[width=1.0\linewidth]{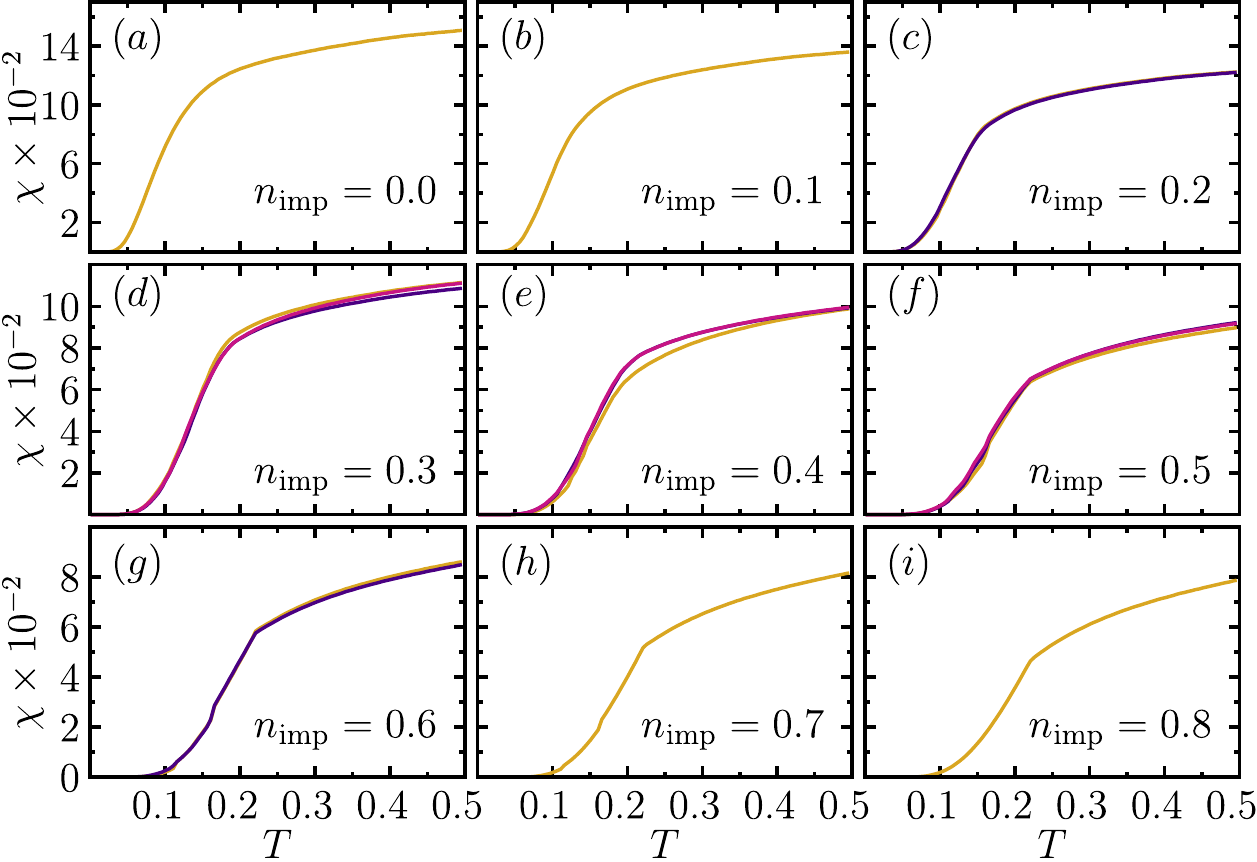}
	\caption{Variations of susceptibility with temperature obtained using CMFT on
	a cluster size consisting of 10 spins ($\Delta_{\rm imp}=2.0$). Dependence on $n_{\rm imp}$
	is shown for various number of random configurations;
	$N_{av}$= 10 (yellow), 20 (blue), and 40 (red).}
	\label{fig:chi_n}
\end{figure}

Presence of impurity bonds induces a N\'eel order, which melts to a disordered state
with increase in temperature. As this effect is more significant in low-temperature regime,
Figures~\ref{fig:Cv_n} and \ref{fig:chi_n} shows the variations of $C_v$ and $\chi$ with
temperature for different $n_{\rm imp}$ with ﬁxed $\Delta_{\rm imp}=2.0$. Transition
temperature increases with increase in $n_{\rm imp}$. For high $n_{\rm imp}$, the peak
in specific heat is very sharp owing to the Ising-like nature of most of the bonds.
For intermediate values of impurity densities, the number of possible random distributions
of the impure spins is huge, leading to difference in magnetic order. Different
colors in Fig.~\ref{fig:Cv_n} correspond to different $N_{av}$. For intermediate values of
$n_{\rm imp}$, specific heat shows a broad peak, which becomes smooth with further
increase in $N_{av}$. Non-monotonic behavior in $C_v$ for intermediate $n_{\rm imp}$
is a direct consequence of the distribution of impure spins. We believe that these peculiarities
might be a finite-size effect and will lead to a clear phase transition in the thermodynamic limit.

Figure~\ref{fig:chi_n} shows dependence of $\chi$ with increasing the number of
random configurations used for averaging. It is interesting to note that susceptibility
decreases with increase in impurity density. Impurity effects in susceptibility
are more prevalent for higher $n_{\rm imp}$, where it shows a discontinuity at $T \sim 0.25$
[see Fig.~\ref{fig:chi_n} (g)-(i)]. Results for susceptibility of SrCo$_2$V$_2$O$_8$~\cite{bera2014magnetic}
show a signature similar to $\chi$ obtained for higher $n_{\rm imp}$. SrCo$_2$V$_2$O$_8$
is expected to be described by XXZ model; however, inter-chain interactions induce
an order in the material at low temperatures. The discontinuity identified in $\chi$ is
a result of vanishing N\`eel order.

\section{Conclusion}

Using DMRG and CMFT, we studied an AFM spin-1/2 Heisenberg chain doped
with spin-$S$ XXZ magnetic impurities, where an XXZ anisotropic impurity
$\Delta_{\rm imp}$ introduces XXZ exchange interaction in the neighboring bonds.
First, we examined the effect of a single spin-$S$ XXZ magnetic impurity in the
ground-state properties such as the spin-spin correlation function, instability of
N\'eel order, and local spin susceptibility. Based on their qualitative characteristics
we find that the types of spin-$S$ impurities are classified into two groups.
$\rm(\hspace{.18em}i\hspace{.18em})$
One contains nonmagnetic and $S=1$ impurities. A short-range AFM correlation
is enhanced but the decay rate of spin-spin correlation function is faster than
that of the undoped spin-1/2 Heisenberg chain. Since they behaves like a
magnetic defect, i.e., vacancy, the global N\'eel order cannot be supported.
$\rm(\hspace{.08em}ii\hspace{.08em})$
The other contains $S=1/2$ and $S>1$ impurities. They hardly change the
decay rate of spin-spin correlation function. However, the N\'eel fluctuation 
is significantly enhanced around the easy-axis XXZ anisotropic impurities
in the spin-1/2 Heisenberg chain, so that long-range N\'eel order can obviously
be stabilized. We also found that the experimentally observed broad NMR spectra
at low temperature in SrCu$_{0.99}$Co$_{0.01}$O$_2$ may be a consequence
of the presence of large $\Delta_{\rm imp}$, $S=3/2$ impurities.

In the latter part of this paper, we focused on the case of $S=1/2$ impurity
as a representative of $\rm(\hspace{.08em}ii\hspace{.08em})$. 
By considering the staggered magnetization $m^z_{\rm st}$ as an order
parameter of the N\'eel state, we confirmed that a finite amount of easy-axis
XXZ $S=1/2$ impurities immediately induces a N\'eel order in the bulk spin
chain and $m^z_{\rm st}$ increases with increasing $\Delta_{\rm imp}$ and
$n_{\rm imp}$. In the presence of uniform magnetic field $h_z$, the total
magnetization $m^z$ exhibits a pseudo-gap behavior at low $h_z$, which is
more pronounced when approaching a pure easy-axis XXZ Heisenberg chain
in the $n_{\rm imp}=1$ limit. Also, a plateau-like feature at $m^z\sim (1-n_{\rm imp})/2$
revealing the existence of isolated magnons is seen at low $n_{\rm imp}$.

Furthermore, we investigated the thermodynamic properties such as
specific heat and magnetic susceptibility using CMFT. These calculations
revealed a phase transition from a N\'eel order to a paramagnet. 
The transition temperature as well as the size of the peak in specific heat
increases with increase in $n_{\rm imp}$ and $\Delta_{\rm imp}$.
The dependence of the transition temperature on the concentration
and anisotropy-strength of impurities is relevant for real systems, where
finite-temperature phase transitions become possible due to inter-chain
couplings.

\section*{Acknowledgements}

We thank Ulrike Nitzsche for technical support. This work was supported by
the SFB 1143 of the Deutsche Forschungsgemeinschaft (Project No. A05)
and by Grants-in-Aid for Scientific Research from JSPS (Projects No. JP17K05530,
No. JP20H01849, and No. JP21J20604). M.K. acknowledges support from
the JSPS Research Fellowship for Young Scientists.

\appendix

\section{Localized $S=1/2$ states near $S_{\rm imp}>1$ impurity}\label{app:threesite}

In Sec.~\ref{sec:lss}, we argue that the $S=1/2$ states are localized around
the $S_{\rm imp}>1$ impurity in the $S^z=1/2$ sector. To explain this, we consider
a simplified three-site Heisenberg system consisting of an $S_{\rm imp}=T/2$ impurity
and the neighboring two $S=1/2$'s. In the $S^z=1/2$ sector, possible bases are
\begin{align}
	\phi_1&=|\frac{1}{2},\frac{1}{2}\rangle \otimes |\frac{T}{2},-\frac{1}{2}\rangle \otimes |\frac{1}{2},\frac{1}{2}\rangle\\
	\phi_2&=|\frac{1}{2},\frac{1}{2}\rangle \otimes |\frac{T}{2},\frac{1}{2}\rangle \otimes |\frac{1}{2},-\frac{1}{2}\rangle\\
	\phi_3&=|\frac{1}{2},-\frac{1}{2}\rangle \otimes |\frac{T}{2},\frac{1}{2}\rangle \otimes |\frac{1}{2},\frac{1}{2}\rangle\\
	\phi_4&=|\frac{1}{2},-\frac{1}{2}\rangle \otimes |\frac{T}{2},\frac{3}{2}\rangle \otimes |\frac{1}{2},-\frac{1}{2}\rangle,
\label{basis}
\end{align}
where $|jm\rangle$ are the simultaneous eigenstates of the angular momentum quantum
number $j$ and the angular momentum projection onto the $z$-axis $m$. For the case of
$\Delta_{\rm imp}=1$, this problem can be easily solved. The ground state energy is
$\varepsilon_0=-(T+2)/2$ and the eigenfunction is
\begin{align}
	\psi_{\rm g.s.}(\Delta_{\rm imp}=1)=A\left(\phi_1-\phi_2-\phi_3+\sqrt{\frac{T+3}{T-1}}\phi_4\right)
		\label{gs_three}
\end{align}
with $A=\sqrt{\frac{T-1}{4T}}$. If $T$ is small enough, the dominant configuration
of ground state may be written as
\begin{align}
	\psi_{\rm g.s.}(\Delta_{\rm imp}=1)\approx|\frac{1}{2},-\frac{1}{2}\rangle \otimes |\frac{T}{2},\frac{3}{2}\rangle \otimes |\frac{1}{2},-\frac{1}{2}\rangle.
	\label{phi_loc_T}
\end{align}
This means that $S^z$ is maximum at the impurity site. Thus, we conclude that
the $S=1/2$ states are localized around the $S_{\rm imp}>1$ impurity.
Since the approximation \eqref{phi_loc_T} approaches the exact ground state
with increasing $\Delta_{\rm imp}$, the localization of $S=1/2$ states is stronger
for larger $\Delta_{\rm imp}$.

In the limit of $\Delta_{\rm imp}=0$, the ground state is
\begin{align}
\nonumber
	&\psi_{\rm g.s.}(\Delta_{\rm imp}=0)=\\&B\left(\frac{T+1}{\sqrt{T^2+2T-1}}\phi_1
	-2\phi_2-2\phi_3+\sqrt{\frac{T^2+2T-3}{T^2+2T-1}}\phi_4\right)
		\label{gs_three_D0}
\end{align}
with $B=1/\sqrt{10}$. The bases (A2) and (A3) have larger coefficients. Thus,
we find that the localization of $S=1/2$ states is released for smaller $\Delta_{\rm imp}$.

\begin{figure}[t]
  \includegraphics[width=0.8\columnwidth]{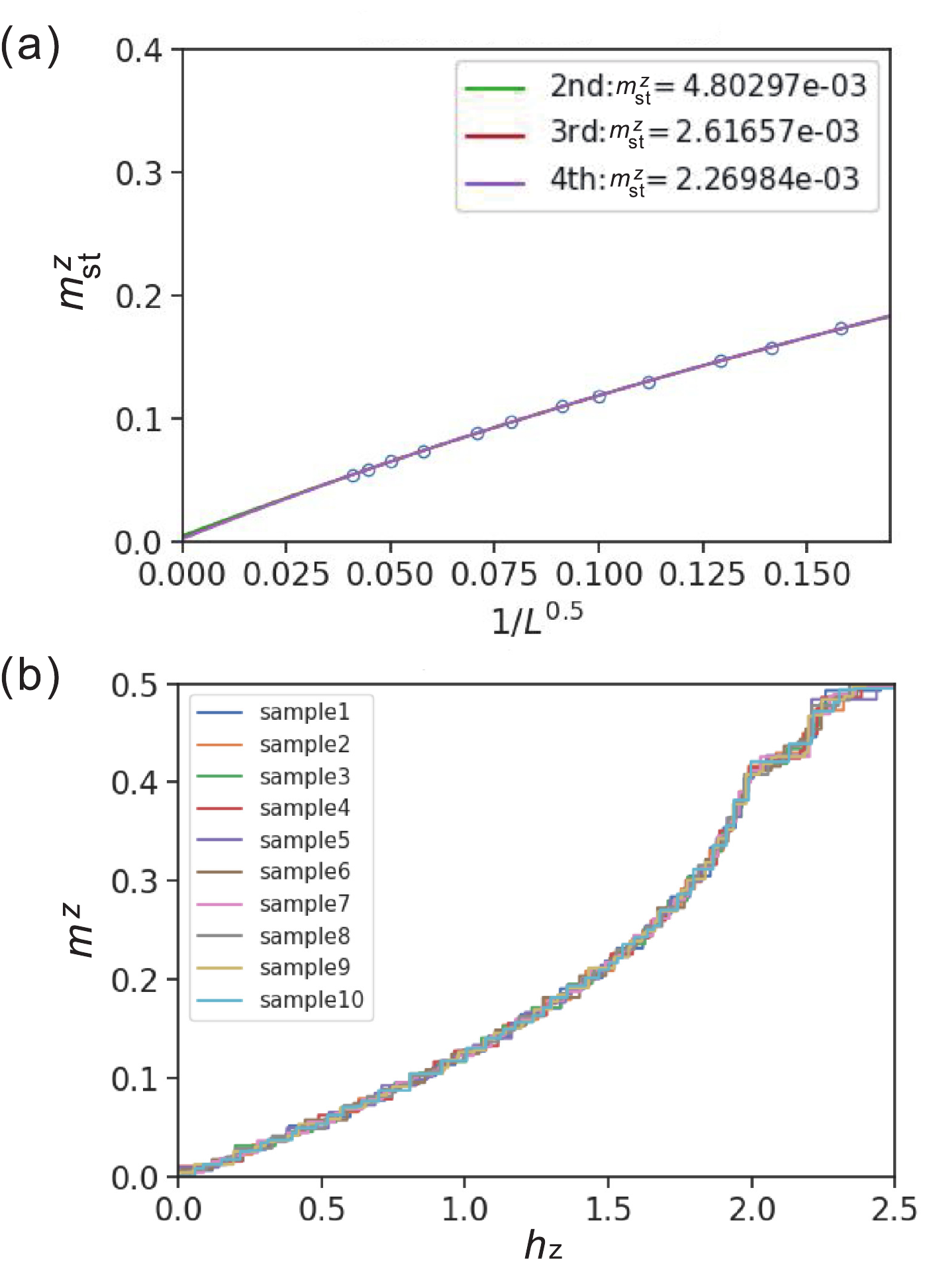}
  \caption{
  (a) Finite-size scaling analysis of the staggered magnetization $m_{\rm st}^z$ calculated
  by DMRG. The parameters are $\Delta_{\rm imp}=1.1$ and $n_{\rm imp} =0.1$.
  Each point is obtained by averaging over $10000/L$ samples of random
  impurity distribution. (b) DMRG results for the uniform magnetization $m^z$ as
  a function of external field. The magnetization curves for 10 random
  samplings of impurity distribution with $L=1000$ are shown.
  }
  \label{fig:scalingDMRG}
\end{figure}

\begin{figure}[tbh]
  \includegraphics[width=0.8\columnwidth]{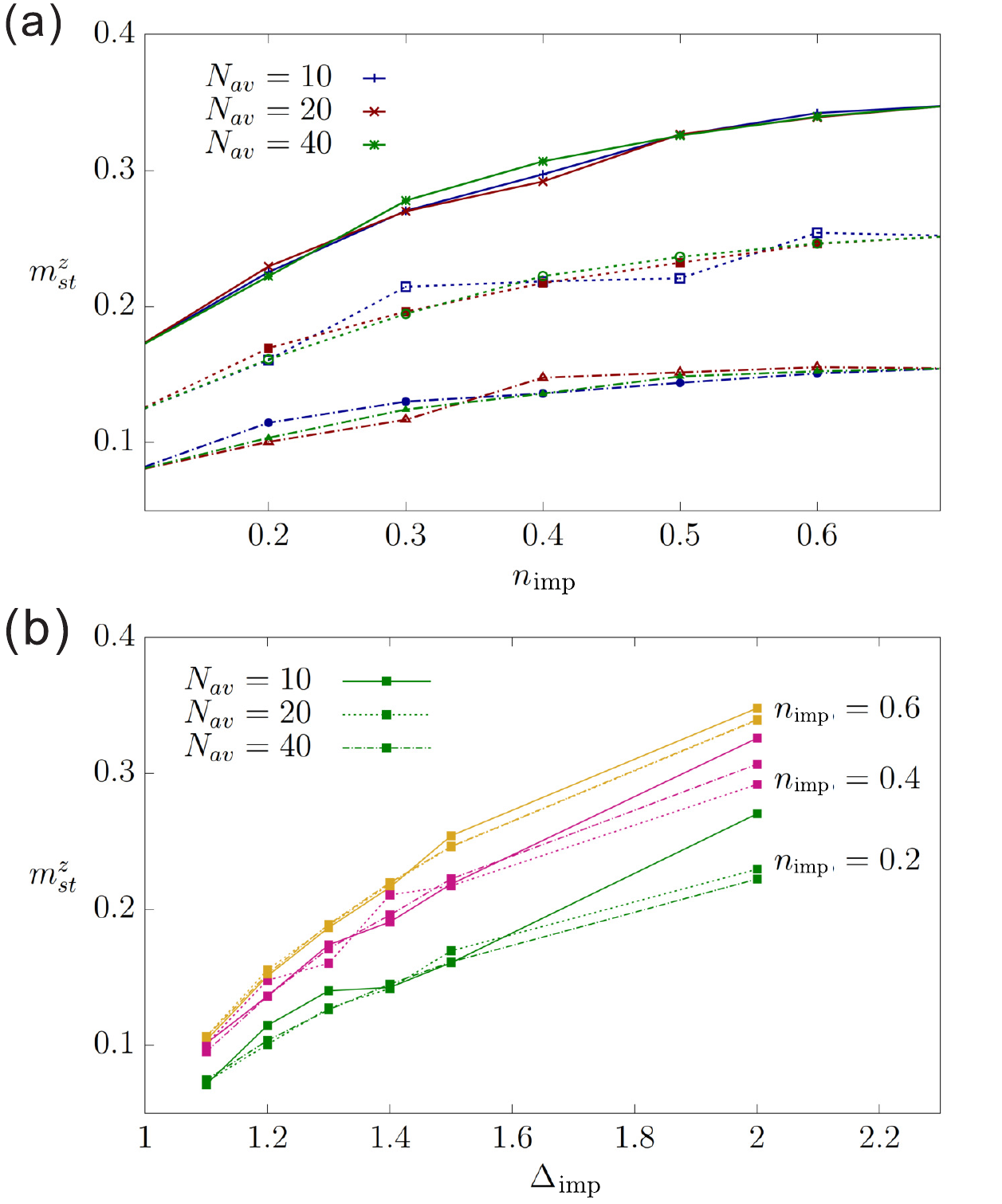}
  \caption{
  (a) Variations of $m^z_{\rm st}$ with $n_{\rm imp}$ for averaging over different number
  of random configurations for $\Delta_{\rm imp}=1.2$ (dot-dashed lines), 
  $\Delta_{\rm imp}=1.5$ (dashed lines),  $\Delta_{\rm imp}=2.0$ (solid lines).
  (b) Variation of $m^z_{\rm st}$ with  $\Delta_{\rm imp}$ with increasing $N_{av}$
  for $n_{\rm imp}=0.2$, $0.4$, and $0.6$.
  }
  \label{fig:sampling_CMFT}
\end{figure}

\section{Finite-size scaling and random sampling in DMRG calculations}\label{app:scalingDMRG}

Figure~\ref{fig:scalingDMRG}(a) shows a finite-scaling analysis of the staggered
magnetization $m^z_{\rm st}$ at $\Delta_{\rm imp}=1.1$ and $n_{\rm imp} =0.1$,
which is one of the most difficult cases to perform the scaling analysis because
of small extrapolated value in the thermodynamic limit $L\to\infty$. Fittings
with the third- and forth-order polynomial functions of $l=\sqrt{L}$ lead to 
$m^z_{\rm st}=2.617\times10^{-3}$ and $m^z_{\rm st}=2.270\times10^{-3}$
in the limit $L\to\infty$, respectively. They are sufficiently close to each other
and the fitting lines seem to be reasonable as shown in Fig.~\ref{fig:scalingDMRG}(a).
Thus, we can confirm the validity of the finite-scaling analysis of $m^z_{\rm st}$.

Figure~\ref{fig:scalingDMRG}(b) shows the uniform magnetization $m^z$ as
a function of external field at $\Delta_{\rm imp}=1.1$ and $n_{\rm imp} =0.1$.
The magnetization curves for 10 random samplings of impurity distribution with
$L=1000$ are plotted. Since we see only small deviations among the
magnetization curves with different random sampling, a system with $L=1000$
may be large enough to reproduce every phenomenon induced by impurity doping.
To gain further accuracy, we average $m^z$ over 10 random samplings
to the magnetization curve shown in the main text.

\section{Sampling of random configurations in CMFT calculations}\label{app:randomsampling}

Staggered magnetization is averaged over various random distribution
of impure bonds for a fixed number of impurities. Figure~\ref{fig:sampling_CMFT}
shows the dependence of $m^z_{\rm st}$ with increasing number of random
configurations considered for averaging. Due to small cluster sizes, the number
of configurations for $n_{\rm imp}=0.1$ and $0.8$ cannot be increased.
It is evident from Fig.~\ref{fig:sampling_CMFT}(a) that the fluctuations appearing
in $m^z_{\rm st}$ can be minimized by increasing the number of configurations.
Figure~\ref{fig:sampling_CMFT}(b) depicts the dependence of the number of
configurations for different $\Delta_{\rm imp}$. It suggests that the dependence
of $m^z_{\rm st}$ on $N_{av}$ is more significant for lower $n_{\rm imp}=0.2$
and $0.4$.

\begin{figure}[tbh]
  \includegraphics[width=0.8\columnwidth]{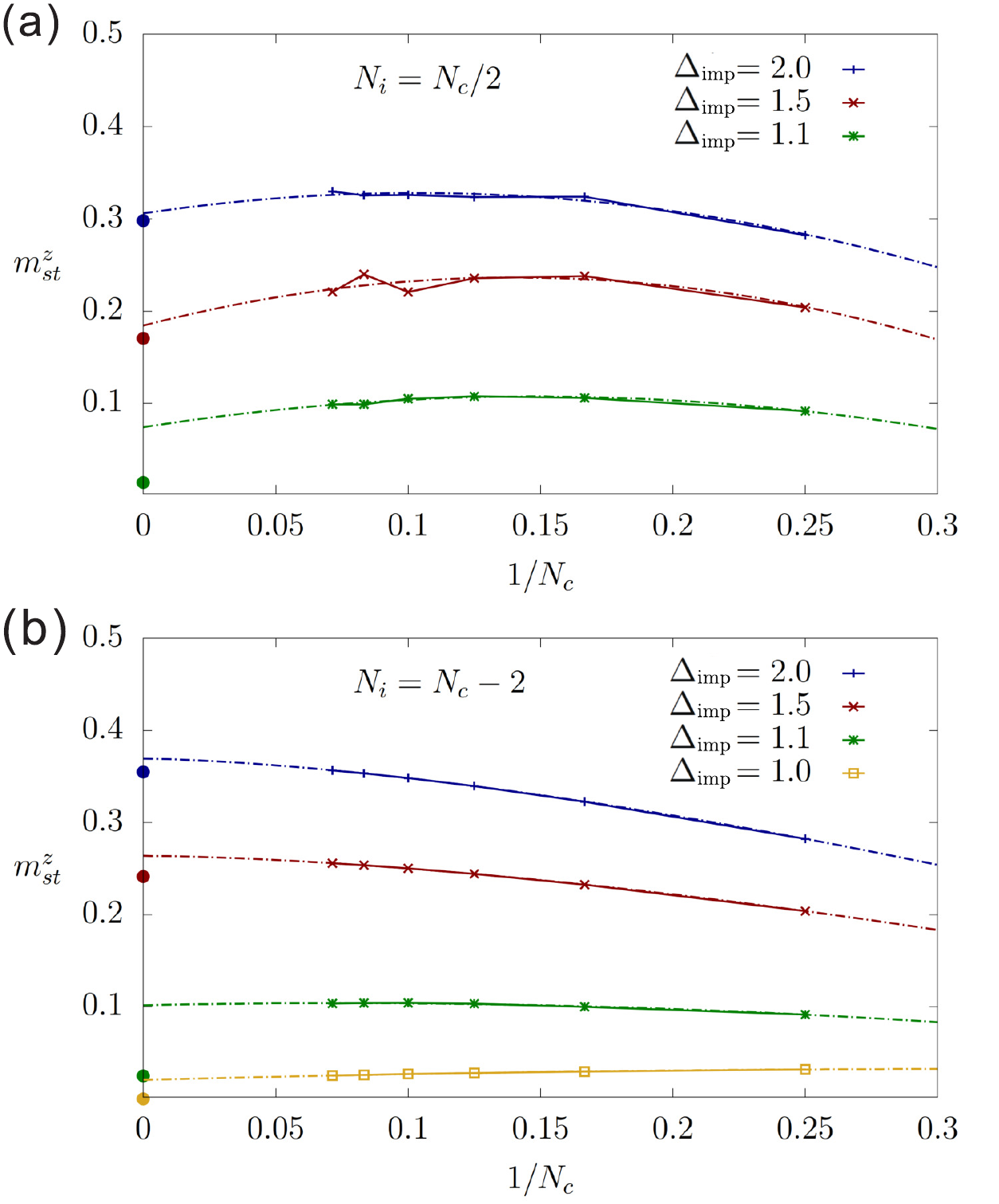}
  \caption{
Cluster-size scaling of $m_{\rm st}^z$ for various $\Delta_{\rm imp}$ with the number
of impurities (a) $N_i=N_c/2$ and (b) $N_i=N_c-2$. Average is taken for the
maximum of 10 random configurations. Dot-dashed lines are power-law fits.
Solid circles are the results computed using DMRG for corresponding
$\Delta_{\rm imp}$ and $n_{\rm imp}$.
  }
  \label{fig:sizeeffect_CMFT}
\end{figure}

\section{Finite-size effects in CMFT calculations}\label{app:sizeeffect}

Figure~\ref{fig:sizeeffect_CMFT} shows $m_{\rm st}^z$ for the number iof
mpurities ($N_i$) fixed to $N_c/2$ and $N_c-2$ for respective cluster sizes.
Fluctuations for the case of $N_i=N_c/2$ appear as the average is taken for
very small number of random configurations ($N_{av}=10$). We observe that
$m_{\rm st}^z$ increases with increasing cluster size up to a certain cluster size,
after which it starts to decrease. Power-law fits to the data suggests that on
further increasing the size, $m_{\rm st}^z$ will reach the value closer to the one
obtained by DMRG [solid circles in Fig.~\ref{fig:sizeeffect_CMFT}(a)].
For $N_i=N_c-2$, finite-size scaling analysis shows an increase in $m_{\rm st}^z$
with increasing $N_c$. This is counter-intuitive as for completely isotropic Ising
($\Delta \to \infty$) or Heisenberg ($\Delta=1$) spin chain, where $m_{\rm st}^z$
decreases with system size. However in our approach, the MF bond remains
Heisenberg-like while the impure bonds are XXZ-type or, in the extreme case,
Ising type. For $N_c=4$, the number of impure bonds within the cluster is 3,
while there are 2 mean-field decoupled Heisenberg bonds. With increasing
number of spins in a cluster, the ratio of impure bonds to pure bonds keeps
on increasing as the number of isotropic MF bonds remain the same. Due to
this competition among Heisenberg bonds and Ising-type (XXZ-type) bonds,
$m_{\rm st}^z$ increases with system size, eventually saturates towards its maximum
possible value. Solid circles in Fig.~\ref{fig:sizeeffect_CMFT} (b) show
the results obtained by DMRG corresponding to $n_{\rm imp}=0.8$.
Power-law ﬁts of CMFT cluster-size scaling show that the agreement with
DMRG results is higher for larger anisotropy in comparison to $\Delta_{\rm imp}=1.1$.

\bibliography{XXZimp}

\begin{thebibliography}{42}%
\makeatletter
\providecommand \@ifxundefined [1]{%
 \@ifx{#1\undefined}
}%
\providecommand \@ifnum [1]{%
 \ifnum #1\expandafter \@firstoftwo
 \else \expandafter \@secondoftwo
 \fi
}%
\providecommand \@ifx [1]{%
 \ifx #1\expandafter \@firstoftwo
 \else \expandafter \@secondoftwo
 \fi
}%
\providecommand \natexlab [1]{#1}%
\providecommand \enquote  [1]{``#1''}%
\providecommand \bibnamefont  [1]{#1}%
\providecommand \bibfnamefont [1]{#1}%
\providecommand \citenamefont [1]{#1}%
\providecommand \href@noop [0]{\@secondoftwo}%
\providecommand \href [0]{\begingroup \@sanitize@url \@href}%
\providecommand \@href[1]{\@@startlink{#1}\@@href}%
\providecommand \@@href[1]{\endgroup#1\@@endlink}%
\providecommand \@sanitize@url [0]{\catcode `\\12\catcode `\$12\catcode
  `\&12\catcode `\#12\catcode `\^12\catcode `\_12\catcode `\%12\relax}%
\providecommand \@@startlink[1]{}%
\providecommand \@@endlink[0]{}%
\providecommand \url  [0]{\begingroup\@sanitize@url \@url }%
\providecommand \@url [1]{\endgroup\@href {#1}{\urlprefix }}%
\providecommand \urlprefix  [0]{URL }%
\providecommand \Eprint [0]{\href }%
\providecommand \doibase [0]{https://doi.org/}%
\providecommand \selectlanguage [0]{\@gobble}%
\providecommand \bibinfo  [0]{\@secondoftwo}%
\providecommand \bibfield  [0]{\@secondoftwo}%
\providecommand \translation [1]{[#1]}%
\providecommand \BibitemOpen [0]{}%
\providecommand \bibitemStop [0]{}%
\providecommand \bibitemNoStop [0]{.\EOS\space}%
\providecommand \EOS [0]{\spacefactor3000\relax}%
\providecommand \BibitemShut  [1]{\csname bibitem#1\endcsname}%
\let\auto@bib@innerbib\@empty
\bibitem [{\citenamefont {Dugdale}(2005)}]{Dugdale2005}%
  \BibitemOpen
  \bibfield  {author} {\bibinfo {author} {\bibfnamefont {J.~S.}\ \bibnamefont
  {Dugdale}},\ }\href@noop {} {\emph {\bibinfo {title} {The electrical
  properties of disordered metals}}}\ (\bibinfo  {publisher} {Cambridge
  University Press},\ \bibinfo {year} {2005})\BibitemShut {NoStop}%
\bibitem [{\citenamefont {Maeda}\ \emph {et~al.}(1990)\citenamefont {Maeda},
  \citenamefont {Yabe}, \citenamefont {Takebayashi}, \citenamefont {Hase},\
  and\ \citenamefont {Uchinokura}}]{Maeda1990}%
  \BibitemOpen
  \bibfield  {author} {\bibinfo {author} {\bibfnamefont {A.}~\bibnamefont
  {Maeda}}, \bibinfo {author} {\bibfnamefont {T.}~\bibnamefont {Yabe}},
  \bibinfo {author} {\bibfnamefont {S.}~\bibnamefont {Takebayashi}}, \bibinfo
  {author} {\bibfnamefont {M.}~\bibnamefont {Hase}},\ and\ \bibinfo {author}
  {\bibfnamefont {K.}~\bibnamefont {Uchinokura}},\ }\bibfield  {title}
  {\bibinfo {title} {Substitution of 3d metals for cu in
  ${\mathrm{bi}}_{2}$(${\mathrm{sr}}_{0.6}$${\mathrm{ca}}_{0.4}$${)}_{3}$${\mathrm{cu}}_{2}$${\mathrm{o}}_{\mathit{y}}$},\
  }\href {https://doi.org/10.1103/PhysRevB.41.4112} {\bibfield  {journal}
  {\bibinfo  {journal} {Phys. Rev. B}\ }\textbf {\bibinfo {volume} {41}},\
  \bibinfo {pages} {4112} (\bibinfo {year} {1990})}\BibitemShut {NoStop}%
\bibitem [{\citenamefont {Anderson}(1958)}]{Anderson1958}%
  \BibitemOpen
  \bibfield  {author} {\bibinfo {author} {\bibfnamefont {P.~W.}\ \bibnamefont
  {Anderson}},\ }\bibfield  {title} {\bibinfo {title} {Absence of diffusion in
  certain random lattices},\ }\href {https://doi.org/10.1103/PhysRev.109.1492}
  {\bibfield  {journal} {\bibinfo  {journal} {Phys. Rev.}\ }\textbf {\bibinfo
  {volume} {109}},\ \bibinfo {pages} {1492} (\bibinfo {year}
  {1958})}\BibitemShut {NoStop}%
\bibitem [{\citenamefont {Ying}\ \emph {et~al.}(2016)\citenamefont {Ying},
  \citenamefont {Gu}, \citenamefont {Chen}, \citenamefont {Wang}, \citenamefont
  {Jin}, \citenamefont {Zhao}, \citenamefont {Zhang},\ and\ \citenamefont
  {Chen}}]{Ying2016}%
  \BibitemOpen
  \bibfield  {author} {\bibinfo {author} {\bibfnamefont {T.}~\bibnamefont
  {Ying}}, \bibinfo {author} {\bibfnamefont {Y.}~\bibnamefont {Gu}}, \bibinfo
  {author} {\bibfnamefont {X.}~\bibnamefont {Chen}}, \bibinfo {author}
  {\bibfnamefont {X.}~\bibnamefont {Wang}}, \bibinfo {author} {\bibfnamefont
  {S.}~\bibnamefont {Jin}}, \bibinfo {author} {\bibfnamefont {L.}~\bibnamefont
  {Zhao}}, \bibinfo {author} {\bibfnamefont {W.}~\bibnamefont {Zhang}},\ and\
  \bibinfo {author} {\bibfnamefont {X.}~\bibnamefont {Chen}},\ }\bibfield
  {title} {\bibinfo {title} {Anderson localization of electrons in single
  crystals: Lixfe7se8},\ }\bibfield  {journal} {\bibinfo  {journal} {Science
  Advances}\ }\textbf {\bibinfo {volume} {2}},\ \href
  {https://doi.org/10.1126/sciadv.1501283} {10.1126/sciadv.1501283} (\bibinfo
  {year} {2016}),\ \Eprint
  {https://arxiv.org/abs/https://advances.sciencemag.org/content/2/2/e1501283.full.pdf}
  {https://advances.sciencemag.org/content/2/2/e1501283.full.pdf} \BibitemShut
  {NoStop}%
\bibitem [{\citenamefont {Imry}\ and\ \citenamefont {Wortis}(1979)}]{Imry1979}%
  \BibitemOpen
  \bibfield  {author} {\bibinfo {author} {\bibfnamefont {Y.}~\bibnamefont
  {Imry}}\ and\ \bibinfo {author} {\bibfnamefont {M.}~\bibnamefont {Wortis}},\
  }\bibfield  {title} {\bibinfo {title} {Influence of quenched impurities on
  first-order phase transitions},\ }\href
  {https://doi.org/10.1103/PhysRevB.19.3580} {\bibfield  {journal} {\bibinfo
  {journal} {Phys. Rev. B}\ }\textbf {\bibinfo {volume} {19}},\ \bibinfo
  {pages} {3580} (\bibinfo {year} {1979})}\BibitemShut {NoStop}%
\bibitem [{\citenamefont {Hammerath}\ \emph {et~al.}(2011)\citenamefont
  {Hammerath}, \citenamefont {Nishimoto}, \citenamefont {Grafe}, \citenamefont
  {Wolter}, \citenamefont {Kataev}, \citenamefont {Ribeiro}, \citenamefont
  {Hess}, \citenamefont {Drechsler},\ and\ \citenamefont
  {B\"uchner}}]{Hammerath2011}%
  \BibitemOpen
  \bibfield  {author} {\bibinfo {author} {\bibfnamefont {F.}~\bibnamefont
  {Hammerath}}, \bibinfo {author} {\bibfnamefont {S.}~\bibnamefont
  {Nishimoto}}, \bibinfo {author} {\bibfnamefont {H.-J.}\ \bibnamefont
  {Grafe}}, \bibinfo {author} {\bibfnamefont {A.~U.~B.}\ \bibnamefont
  {Wolter}}, \bibinfo {author} {\bibfnamefont {V.}~\bibnamefont {Kataev}},
  \bibinfo {author} {\bibfnamefont {P.}~\bibnamefont {Ribeiro}}, \bibinfo
  {author} {\bibfnamefont {C.}~\bibnamefont {Hess}}, \bibinfo {author}
  {\bibfnamefont {S.-L.}\ \bibnamefont {Drechsler}},\ and\ \bibinfo {author}
  {\bibfnamefont {B.}~\bibnamefont {B\"uchner}},\ }\bibfield  {title} {\bibinfo
  {title} {Spin gap in the zigzag spin-$1/2$ chain cuprate
  ${\mathrm{sr}}_{0.9}{\mathrm{ca}}_{0.1}{\mathrm{cuo}}_{2}$},\ }\href
  {https://doi.org/10.1103/PhysRevLett.107.017203} {\bibfield  {journal}
  {\bibinfo  {journal} {Phys. Rev. Lett.}\ }\textbf {\bibinfo {volume} {107}},\
  \bibinfo {pages} {017203} (\bibinfo {year} {2011})}\BibitemShut {NoStop}%
\bibitem [{\citenamefont {Eggert}\ and\ \citenamefont
  {Affleck}(1992)}]{Eggert1992}%
  \BibitemOpen
  \bibfield  {author} {\bibinfo {author} {\bibfnamefont {S.}~\bibnamefont
  {Eggert}}\ and\ \bibinfo {author} {\bibfnamefont {I.}~\bibnamefont
  {Affleck}},\ }\bibfield  {title} {\bibinfo {title} {Magnetic impurities in
  half-integer-spin heisenberg antiferromagnetic chains},\ }\href
  {https://doi.org/10.1103/PhysRevB.46.10866} {\bibfield  {journal} {\bibinfo
  {journal} {Phys. Rev. B}\ }\textbf {\bibinfo {volume} {46}},\ \bibinfo
  {pages} {10866} (\bibinfo {year} {1992})}\BibitemShut {NoStop}%
\bibitem [{\citenamefont {Anfuso}\ and\ \citenamefont
  {Eggert}(2006)}]{Anfuso2006}%
  \BibitemOpen
  \bibfield  {author} {\bibinfo {author} {\bibfnamefont {F.}~\bibnamefont
  {Anfuso}}\ and\ \bibinfo {author} {\bibfnamefont {S.}~\bibnamefont
  {Eggert}},\ }\bibfield  {title} {\bibinfo {title} {Interaction effects
  between impurities in low-dimensional spin-(1/2) antiferromagnets},\ }\href
  {https://doi.org/10.1209/epl/i2005-10380-y} {\bibfield  {journal} {\bibinfo
  {journal} {Europhysics Letters ({EPL})}\ }\textbf {\bibinfo {volume} {73}},\
  \bibinfo {pages} {271} (\bibinfo {year} {2006})}\BibitemShut {NoStop}%
\bibitem [{\citenamefont {Hagiwara}\ \emph {et~al.}(1990)\citenamefont
  {Hagiwara}, \citenamefont {Katsumata}, \citenamefont {Affleck}, \citenamefont
  {Halperin},\ and\ \citenamefont {Renard}}]{Hagiwara1990}%
  \BibitemOpen
  \bibfield  {author} {\bibinfo {author} {\bibfnamefont {M.}~\bibnamefont
  {Hagiwara}}, \bibinfo {author} {\bibfnamefont {K.}~\bibnamefont {Katsumata}},
  \bibinfo {author} {\bibfnamefont {I.}~\bibnamefont {Affleck}}, \bibinfo
  {author} {\bibfnamefont {B.~I.}\ \bibnamefont {Halperin}},\ and\ \bibinfo
  {author} {\bibfnamefont {J.~P.}\ \bibnamefont {Renard}},\ }\bibfield  {title}
  {\bibinfo {title} {Observation of s=1/2 degrees of freedom in an s=1
  linear-chain heisenberg antiferromagnet},\ }\href
  {https://doi.org/10.1103/PhysRevLett.65.3181} {\bibfield  {journal} {\bibinfo
   {journal} {Phys. Rev. Lett.}\ }\textbf {\bibinfo {volume} {65}},\ \bibinfo
  {pages} {3181} (\bibinfo {year} {1990})}\BibitemShut {NoStop}%
\bibitem [{\citenamefont {Hase}\ \emph {et~al.}(1993)\citenamefont {Hase},
  \citenamefont {Terasaki}, \citenamefont {Sasago}, \citenamefont
  {Uchinokura},\ and\ \citenamefont {Obara}}]{Hase1993}%
  \BibitemOpen
  \bibfield  {author} {\bibinfo {author} {\bibfnamefont {M.}~\bibnamefont
  {Hase}}, \bibinfo {author} {\bibfnamefont {I.}~\bibnamefont {Terasaki}},
  \bibinfo {author} {\bibfnamefont {Y.}~\bibnamefont {Sasago}}, \bibinfo
  {author} {\bibfnamefont {K.}~\bibnamefont {Uchinokura}},\ and\ \bibinfo
  {author} {\bibfnamefont {H.}~\bibnamefont {Obara}},\ }\bibfield  {title}
  {\bibinfo {title} {Effects of substitution of zn for cu in the spin-peierls
  cuprate, ${\mathrm{cugeo}}_{3}$: The suppression of the spin-peierls
  transition and the occurrence of a new spin-glass state},\ }\href
  {https://doi.org/10.1103/PhysRevLett.71.4059} {\bibfield  {journal} {\bibinfo
   {journal} {Phys. Rev. Lett.}\ }\textbf {\bibinfo {volume} {71}},\ \bibinfo
  {pages} {4059} (\bibinfo {year} {1993})}\BibitemShut {NoStop}%
\bibitem [{\citenamefont {Grenier}\ \emph {et~al.}(1997)\citenamefont
  {Grenier}, \citenamefont {Regnault}, \citenamefont {Lorenzo}, \citenamefont
  {Bossy}, \citenamefont {Renard}, \citenamefont {Dhalenne},\ and\
  \citenamefont {Revcolevschi}}]{Grenier1997}%
  \BibitemOpen
  \bibfield  {author} {\bibinfo {author} {\bibfnamefont {B.}~\bibnamefont
  {Grenier}}, \bibinfo {author} {\bibfnamefont {L.}~\bibnamefont {Regnault}},
  \bibinfo {author} {\bibfnamefont {J.}~\bibnamefont {Lorenzo}}, \bibinfo
  {author} {\bibfnamefont {J.}~\bibnamefont {Bossy}}, \bibinfo {author}
  {\bibfnamefont {J.}~\bibnamefont {Renard}}, \bibinfo {author} {\bibfnamefont
  {G.}~\bibnamefont {Dhalenne}},\ and\ \bibinfo {author} {\bibfnamefont
  {A.}~\bibnamefont {Revcolevschi}},\ }\bibfield  {title} {\bibinfo {title}
  {Effect of magnetic field and si-doping on the spin-peierls phase of
  cugeo3},\ }\href
  {https://doi.org/https://doi.org/10.1016/S0921-4526(96)01058-7} {\bibfield
  {journal} {\bibinfo  {journal} {Physica B: Condensed Matter}\ }\textbf
  {\bibinfo {volume} {234-236}},\ \bibinfo {pages} {534} (\bibinfo {year}
  {1997})},\ \bibinfo {note} {proceedings of the First European Conference on
  Neutron Scattering}\BibitemShut {NoStop}%
\bibitem [{\citenamefont {Azuma}\ \emph {et~al.}(1997)\citenamefont {Azuma},
  \citenamefont {Fujishiro}, \citenamefont {Takano}, \citenamefont {Nohara},\
  and\ \citenamefont {Takagi}}]{Azuma1997}%
  \BibitemOpen
  \bibfield  {author} {\bibinfo {author} {\bibfnamefont {M.}~\bibnamefont
  {Azuma}}, \bibinfo {author} {\bibfnamefont {Y.}~\bibnamefont {Fujishiro}},
  \bibinfo {author} {\bibfnamefont {M.}~\bibnamefont {Takano}}, \bibinfo
  {author} {\bibfnamefont {M.}~\bibnamefont {Nohara}},\ and\ \bibinfo {author}
  {\bibfnamefont {H.}~\bibnamefont {Takagi}},\ }\bibfield  {title} {\bibinfo
  {title} {Switching of the gapped singlet spin-liquid state to an
  antiferromagnetically ordered state in
  sr(${\mathrm{cu}}_{1\mathrm{\ensuremath{-}}\mathrm{x}}$${\mathrm{zn}}_{\mathrm{x}}$${)}_{2}$${\mathrm{o}}_{3}$},\
  }\href {https://doi.org/10.1103/PhysRevB.55.R8658} {\bibfield  {journal}
  {\bibinfo  {journal} {Phys. Rev. B}\ }\textbf {\bibinfo {volume} {55}},\
  \bibinfo {pages} {R8658} (\bibinfo {year} {1997})}\BibitemShut {NoStop}%
\bibitem [{\citenamefont {Motoyama}\ \emph {et~al.}(1996)\citenamefont
  {Motoyama}, \citenamefont {Eisaki},\ and\ \citenamefont
  {Uchida}}]{Motoyama1996}%
  \BibitemOpen
  \bibfield  {author} {\bibinfo {author} {\bibfnamefont {N.}~\bibnamefont
  {Motoyama}}, \bibinfo {author} {\bibfnamefont {H.}~\bibnamefont {Eisaki}},\
  and\ \bibinfo {author} {\bibfnamefont {S.}~\bibnamefont {Uchida}},\
  }\bibfield  {title} {\bibinfo {title} {Magnetic susceptibility of ideal spin
  1 $/$2 heisenberg antiferromagnetic chain systems,
  ${\mathrm{sr}}_{2}{\mathrm{cuo}}_{3}$ and ${\mathrm{srcuo}}_{2}$},\ }\href
  {https://doi.org/10.1103/PhysRevLett.76.3212} {\bibfield  {journal} {\bibinfo
   {journal} {Phys. Rev. Lett.}\ }\textbf {\bibinfo {volume} {76}},\ \bibinfo
  {pages} {3212} (\bibinfo {year} {1996})}\BibitemShut {NoStop}%
\bibitem [{\citenamefont {Kojima}\ \emph {et~al.}(2004)\citenamefont {Kojima},
  \citenamefont {Yamanobe}, \citenamefont {Eisaki}, \citenamefont {Uchida},
  \citenamefont {Fudamoto}, \citenamefont {Gat}, \citenamefont {Larkin},
  \citenamefont {Savici}, \citenamefont {Uemura}, \citenamefont {Kyriakou},
  \citenamefont {Rovers},\ and\ \citenamefont {Luke}}]{Kojima2004}%
  \BibitemOpen
  \bibfield  {author} {\bibinfo {author} {\bibfnamefont {K.~M.}\ \bibnamefont
  {Kojima}}, \bibinfo {author} {\bibfnamefont {J.}~\bibnamefont {Yamanobe}},
  \bibinfo {author} {\bibfnamefont {H.}~\bibnamefont {Eisaki}}, \bibinfo
  {author} {\bibfnamefont {S.}~\bibnamefont {Uchida}}, \bibinfo {author}
  {\bibfnamefont {Y.}~\bibnamefont {Fudamoto}}, \bibinfo {author}
  {\bibfnamefont {I.~M.}\ \bibnamefont {Gat}}, \bibinfo {author} {\bibfnamefont
  {M.~I.}\ \bibnamefont {Larkin}}, \bibinfo {author} {\bibfnamefont
  {A.}~\bibnamefont {Savici}}, \bibinfo {author} {\bibfnamefont {Y.~J.}\
  \bibnamefont {Uemura}}, \bibinfo {author} {\bibfnamefont {P.~P.}\
  \bibnamefont {Kyriakou}}, \bibinfo {author} {\bibfnamefont {M.~T.}\
  \bibnamefont {Rovers}},\ and\ \bibinfo {author} {\bibfnamefont {G.~M.}\
  \bibnamefont {Luke}},\ }\bibfield  {title} {\bibinfo {title} {Site-dilution
  in the quasi-one-dimensional antiferromagnet
  ${\mathrm{sr}}_{2}({\mathrm{cu}}_{1\ensuremath{-}x}{\mathrm{pd}}_{x}){\mathrm{o}}_{3}$:
  Reduction of n\'eel temperature and spatial distribution of ordered moment
  sizes},\ }\href {https://doi.org/10.1103/PhysRevB.70.094402} {\bibfield
  {journal} {\bibinfo  {journal} {Phys. Rev. B}\ }\textbf {\bibinfo {volume}
  {70}},\ \bibinfo {pages} {094402} (\bibinfo {year} {2004})}\BibitemShut
  {NoStop}%
\bibitem [{\citenamefont {Simutis}\ \emph {et~al.}(2013)\citenamefont
  {Simutis}, \citenamefont {Gvasaliya}, \citenamefont {M\aa{}nsson},
  \citenamefont {Chernyshev}, \citenamefont {Mohan}, \citenamefont {Singh},
  \citenamefont {Hess}, \citenamefont {Savici}, \citenamefont {Kolesnikov},
  \citenamefont {Piovano}, \citenamefont {Perring}, \citenamefont {Zaliznyak},
  \citenamefont {B\"uchner},\ and\ \citenamefont {Zheludev}}]{Simutis2013}%
  \BibitemOpen
  \bibfield  {author} {\bibinfo {author} {\bibfnamefont {G.}~\bibnamefont
  {Simutis}}, \bibinfo {author} {\bibfnamefont {S.}~\bibnamefont {Gvasaliya}},
  \bibinfo {author} {\bibfnamefont {M.}~\bibnamefont {M\aa{}nsson}}, \bibinfo
  {author} {\bibfnamefont {A.~L.}\ \bibnamefont {Chernyshev}}, \bibinfo
  {author} {\bibfnamefont {A.}~\bibnamefont {Mohan}}, \bibinfo {author}
  {\bibfnamefont {S.}~\bibnamefont {Singh}}, \bibinfo {author} {\bibfnamefont
  {C.}~\bibnamefont {Hess}}, \bibinfo {author} {\bibfnamefont {A.~T.}\
  \bibnamefont {Savici}}, \bibinfo {author} {\bibfnamefont {A.~I.}\
  \bibnamefont {Kolesnikov}}, \bibinfo {author} {\bibfnamefont
  {A.}~\bibnamefont {Piovano}}, \bibinfo {author} {\bibfnamefont
  {T.}~\bibnamefont {Perring}}, \bibinfo {author} {\bibfnamefont
  {I.}~\bibnamefont {Zaliznyak}}, \bibinfo {author} {\bibfnamefont
  {B.}~\bibnamefont {B\"uchner}},\ and\ \bibinfo {author} {\bibfnamefont
  {A.}~\bibnamefont {Zheludev}},\ }\bibfield  {title} {\bibinfo {title} {Spin
  pseudogap in ni-doped ${\mathrm{srcuo}}_{2}$},\ }\href
  {https://doi.org/10.1103/PhysRevLett.111.067204} {\bibfield  {journal}
  {\bibinfo  {journal} {Phys. Rev. Lett.}\ }\textbf {\bibinfo {volume} {111}},\
  \bibinfo {pages} {067204} (\bibinfo {year} {2013})}\BibitemShut {NoStop}%
\bibitem [{\citenamefont {Karmakar}\ \emph {et~al.}(2017)\citenamefont
  {Karmakar}, \citenamefont {Bag}, \citenamefont {Skoulatos}, \citenamefont
  {R\"uegg},\ and\ \citenamefont {Singh}}]{Karmakar2017}%
  \BibitemOpen
  \bibfield  {author} {\bibinfo {author} {\bibfnamefont {K.}~\bibnamefont
  {Karmakar}}, \bibinfo {author} {\bibfnamefont {R.}~\bibnamefont {Bag}},
  \bibinfo {author} {\bibfnamefont {M.}~\bibnamefont {Skoulatos}}, \bibinfo
  {author} {\bibfnamefont {C.}~\bibnamefont {R\"uegg}},\ and\ \bibinfo {author}
  {\bibfnamefont {S.}~\bibnamefont {Singh}},\ }\bibfield  {title} {\bibinfo
  {title} {Impurities in the weakly coupled quantum spin chains
  ${\mathbf{sr}}_{2}{\mathbf{cuo}}_{3}$ and ${\mathbf{srcuo}}_{2}$},\ }\href
  {https://doi.org/10.1103/PhysRevB.95.235154} {\bibfield  {journal} {\bibinfo
  {journal} {Phys. Rev. B}\ }\textbf {\bibinfo {volume} {95}},\ \bibinfo
  {pages} {235154} (\bibinfo {year} {2017})}\BibitemShut {NoStop}%
\bibitem [{\citenamefont {Kimura}\ \emph
  {et~al.}(2008{\natexlab{a}})\citenamefont {Kimura}, \citenamefont {Matsuda},
  \citenamefont {Masuda}, \citenamefont {Hondo}, \citenamefont {Kaneko},
  \citenamefont {Metoki}, \citenamefont {Hagiwara}, \citenamefont {Takeuchi},
  \citenamefont {Okunishi}, \citenamefont {He}, \citenamefont {Kindo},
  \citenamefont {Taniyama},\ and\ \citenamefont {Itoh}}]{Kimura2008}%
  \BibitemOpen
  \bibfield  {author} {\bibinfo {author} {\bibfnamefont {S.}~\bibnamefont
  {Kimura}}, \bibinfo {author} {\bibfnamefont {M.}~\bibnamefont {Matsuda}},
  \bibinfo {author} {\bibfnamefont {T.}~\bibnamefont {Masuda}}, \bibinfo
  {author} {\bibfnamefont {S.}~\bibnamefont {Hondo}}, \bibinfo {author}
  {\bibfnamefont {K.}~\bibnamefont {Kaneko}}, \bibinfo {author} {\bibfnamefont
  {N.}~\bibnamefont {Metoki}}, \bibinfo {author} {\bibfnamefont
  {M.}~\bibnamefont {Hagiwara}}, \bibinfo {author} {\bibfnamefont
  {T.}~\bibnamefont {Takeuchi}}, \bibinfo {author} {\bibfnamefont
  {K.}~\bibnamefont {Okunishi}}, \bibinfo {author} {\bibfnamefont
  {Z.}~\bibnamefont {He}}, \bibinfo {author} {\bibfnamefont {K.}~\bibnamefont
  {Kindo}}, \bibinfo {author} {\bibfnamefont {T.}~\bibnamefont {Taniyama}},\
  and\ \bibinfo {author} {\bibfnamefont {M.}~\bibnamefont {Itoh}},\ }\bibfield
  {title} {\bibinfo {title} {Longitudinal spin density wave order in a quasi-1d
  ising-like quantum antiferromagnet},\ }\href
  {https://doi.org/10.1103/PhysRevLett.101.207201} {\bibfield  {journal}
  {\bibinfo  {journal} {Phys. Rev. Lett.}\ }\textbf {\bibinfo {volume} {101}},\
  \bibinfo {pages} {207201} (\bibinfo {year} {2008}{\natexlab{a}})}\BibitemShut
  {NoStop}%
\bibitem [{\citenamefont {Bera}\ \emph
  {et~al.}(2014{\natexlab{a}})\citenamefont {Bera}, \citenamefont {Lake},
  \citenamefont {Stein},\ and\ \citenamefont {Zander}}]{Bera2014}%
  \BibitemOpen
  \bibfield  {author} {\bibinfo {author} {\bibfnamefont {A.~K.}\ \bibnamefont
  {Bera}}, \bibinfo {author} {\bibfnamefont {B.}~\bibnamefont {Lake}}, \bibinfo
  {author} {\bibfnamefont {W.-D.}\ \bibnamefont {Stein}},\ and\ \bibinfo
  {author} {\bibfnamefont {S.}~\bibnamefont {Zander}},\ }\bibfield  {title}
  {\bibinfo {title} {Magnetic correlations of the quasi-one-dimensional
  half-integer spin-chain antiferromagnets sr${M}_{2}$v${}_{2}$o${}_{8}$ ($m$ =
  co, mn)},\ }\href {https://doi.org/10.1103/PhysRevB.89.094402} {\bibfield
  {journal} {\bibinfo  {journal} {Phys. Rev. B}\ }\textbf {\bibinfo {volume}
  {89}},\ \bibinfo {pages} {094402} (\bibinfo {year}
  {2014}{\natexlab{a}})}\BibitemShut {NoStop}%
\bibitem [{\citenamefont {Utz}\ \emph {et~al.}(2017)\citenamefont {Utz},
  \citenamefont {Hammerath}, \citenamefont {Kraus}, \citenamefont {Ritschel},
  \citenamefont {Geck}, \citenamefont {Hozoi}, \citenamefont {van~den Brink},
  \citenamefont {Mohan}, \citenamefont {Hess}, \citenamefont {Karmakar},
  \citenamefont {Singh}, \citenamefont {Bounoua}, \citenamefont {Saint-Martin},
  \citenamefont {Pinsard-Gaudart}, \citenamefont {Revcolevschi}, \citenamefont
  {B\"uchner},\ and\ \citenamefont {Grafe}}]{Utz2017}%
  \BibitemOpen
  \bibfield  {author} {\bibinfo {author} {\bibfnamefont {Y.}~\bibnamefont
  {Utz}}, \bibinfo {author} {\bibfnamefont {F.}~\bibnamefont {Hammerath}},
  \bibinfo {author} {\bibfnamefont {R.}~\bibnamefont {Kraus}}, \bibinfo
  {author} {\bibfnamefont {T.}~\bibnamefont {Ritschel}}, \bibinfo {author}
  {\bibfnamefont {J.}~\bibnamefont {Geck}}, \bibinfo {author} {\bibfnamefont
  {L.}~\bibnamefont {Hozoi}}, \bibinfo {author} {\bibfnamefont
  {J.}~\bibnamefont {van~den Brink}}, \bibinfo {author} {\bibfnamefont
  {A.}~\bibnamefont {Mohan}}, \bibinfo {author} {\bibfnamefont
  {C.}~\bibnamefont {Hess}}, \bibinfo {author} {\bibfnamefont {K.}~\bibnamefont
  {Karmakar}}, \bibinfo {author} {\bibfnamefont {S.}~\bibnamefont {Singh}},
  \bibinfo {author} {\bibfnamefont {D.}~\bibnamefont {Bounoua}}, \bibinfo
  {author} {\bibfnamefont {R.}~\bibnamefont {Saint-Martin}}, \bibinfo {author}
  {\bibfnamefont {L.}~\bibnamefont {Pinsard-Gaudart}}, \bibinfo {author}
  {\bibfnamefont {A.}~\bibnamefont {Revcolevschi}}, \bibinfo {author}
  {\bibfnamefont {B.}~\bibnamefont {B\"uchner}},\ and\ \bibinfo {author}
  {\bibfnamefont {H.-J.}\ \bibnamefont {Grafe}},\ }\bibfield  {title} {\bibinfo
  {title} {Effect of different in-chain impurities on the magnetic properties
  of the spin chain compound ${\mathrm{srcuo}}_{2}$ probed by nmr},\ }\href
  {https://doi.org/10.1103/PhysRevB.96.115135} {\bibfield  {journal} {\bibinfo
  {journal} {Phys. Rev. B}\ }\textbf {\bibinfo {volume} {96}},\ \bibinfo
  {pages} {115135} (\bibinfo {year} {2017})}\BibitemShut {NoStop}%
\bibitem [{\citenamefont {Eggert}\ \emph {et~al.}(2001)\citenamefont {Eggert},
  \citenamefont {Gustafsson},\ and\ \citenamefont {Rommer}}]{Eggert2001}%
  \BibitemOpen
  \bibfield  {author} {\bibinfo {author} {\bibfnamefont {S.}~\bibnamefont
  {Eggert}}, \bibinfo {author} {\bibfnamefont {D.~P.}\ \bibnamefont
  {Gustafsson}},\ and\ \bibinfo {author} {\bibfnamefont {S.}~\bibnamefont
  {Rommer}},\ }\bibfield  {title} {\bibinfo {title} {Phase diagram of an
  impurity in the spin- $1/2$ chain: Two-channel kondo effect versus curie
  law},\ }\href {https://doi.org/10.1103/PhysRevLett.86.516} {\bibfield
  {journal} {\bibinfo  {journal} {Phys. Rev. Lett.}\ }\textbf {\bibinfo
  {volume} {86}},\ \bibinfo {pages} {516} (\bibinfo {year} {2001})}\BibitemShut
  {NoStop}%
\bibitem [{\citenamefont {Bobroff}\ \emph {et~al.}(2009)\citenamefont
  {Bobroff}, \citenamefont {Laflorencie}, \citenamefont {Alexander},
  \citenamefont {Mahajan}, \citenamefont {Koteswararao},\ and\ \citenamefont
  {Mendels}}]{Bobroff2009}%
  \BibitemOpen
  \bibfield  {author} {\bibinfo {author} {\bibfnamefont {J.}~\bibnamefont
  {Bobroff}}, \bibinfo {author} {\bibfnamefont {N.}~\bibnamefont
  {Laflorencie}}, \bibinfo {author} {\bibfnamefont {L.~K.}\ \bibnamefont
  {Alexander}}, \bibinfo {author} {\bibfnamefont {A.~V.}\ \bibnamefont
  {Mahajan}}, \bibinfo {author} {\bibfnamefont {B.}~\bibnamefont
  {Koteswararao}},\ and\ \bibinfo {author} {\bibfnamefont {P.}~\bibnamefont
  {Mendels}},\ }\bibfield  {title} {\bibinfo {title} {Impurity-induced magnetic
  order in low-dimensional spin-gapped materials},\ }\href
  {https://doi.org/10.1103/PhysRevLett.103.047201} {\bibfield  {journal}
  {\bibinfo  {journal} {Phys. Rev. Lett.}\ }\textbf {\bibinfo {volume} {103}},\
  \bibinfo {pages} {047201} (\bibinfo {year} {2009})}\BibitemShut {NoStop}%
\bibitem [{\citenamefont {Alexander}\ \emph {et~al.}(2010)\citenamefont
  {Alexander}, \citenamefont {Bobroff}, \citenamefont {Mahajan}, \citenamefont
  {Koteswararao}, \citenamefont {Laflorencie},\ and\ \citenamefont
  {Alet}}]{Alexander2010}%
  \BibitemOpen
  \bibfield  {author} {\bibinfo {author} {\bibfnamefont {L.~K.}\ \bibnamefont
  {Alexander}}, \bibinfo {author} {\bibfnamefont {J.}~\bibnamefont {Bobroff}},
  \bibinfo {author} {\bibfnamefont {A.~V.}\ \bibnamefont {Mahajan}}, \bibinfo
  {author} {\bibfnamefont {B.}~\bibnamefont {Koteswararao}}, \bibinfo {author}
  {\bibfnamefont {N.}~\bibnamefont {Laflorencie}},\ and\ \bibinfo {author}
  {\bibfnamefont {F.}~\bibnamefont {Alet}},\ }\bibfield  {title} {\bibinfo
  {title} {Impurity effects in coupled-ladder
  ${\text{bicu}}_{2}{\text{po}}_{6}$ studied by nmr and quantum monte carlo
  simulations},\ }\href {https://doi.org/10.1103/PhysRevB.81.054438} {\bibfield
   {journal} {\bibinfo  {journal} {Phys. Rev. B}\ }\textbf {\bibinfo {volume}
  {81}},\ \bibinfo {pages} {054438} (\bibinfo {year} {2010})}\BibitemShut
  {NoStop}%
\bibitem [{\citenamefont {Zhang}\ \emph
  {et~al.}(1997{\natexlab{a}})\citenamefont {Zhang}, \citenamefont {Igarashi},\
  and\ \citenamefont {Fulde}}]{Zhang1997}%
  \BibitemOpen
  \bibfield  {author} {\bibinfo {author} {\bibfnamefont {W.}~\bibnamefont
  {Zhang}}, \bibinfo {author} {\bibfnamefont {J.-i.}\ \bibnamefont
  {Igarashi}},\ and\ \bibinfo {author} {\bibfnamefont {P.}~\bibnamefont
  {Fulde}},\ }\bibfield  {title} {\bibinfo {title} {Thermodynamics of an
  impurity coupled to a heisenberg chain: Density-matrix renormalization group
  and monte carlo studies},\ }\href {https://doi.org/10.1143/JPSJ.66.1912}
  {\bibfield  {journal} {\bibinfo  {journal} {Journal of the Physical Society
  of Japan}\ }\textbf {\bibinfo {volume} {66}},\ \bibinfo {pages} {1912}
  (\bibinfo {year} {1997}{\natexlab{a}})},\ \Eprint
  {https://arxiv.org/abs/https://doi.org/10.1143/JPSJ.66.1912}
  {https://doi.org/10.1143/JPSJ.66.1912} \BibitemShut {NoStop}%
\bibitem [{\citenamefont {Zhang}\ \emph
  {et~al.}(1997{\natexlab{b}})\citenamefont {Zhang}, \citenamefont {Igarashi},\
  and\ \citenamefont {Fulde}}]{Zhang1997-2}%
  \BibitemOpen
  \bibfield  {author} {\bibinfo {author} {\bibfnamefont {W.}~\bibnamefont
  {Zhang}}, \bibinfo {author} {\bibfnamefont {J.}~\bibnamefont {Igarashi}},\
  and\ \bibinfo {author} {\bibfnamefont {P.}~\bibnamefont {Fulde}},\ }\bibfield
   {title} {\bibinfo {title} {Magnetic impurity coupled to a heisenberg chain:
  Density-matrix renormalization-group study},\ }\href
  {https://doi.org/10.1103/PhysRevB.56.654} {\bibfield  {journal} {\bibinfo
  {journal} {Phys. Rev. B}\ }\textbf {\bibinfo {volume} {56}},\ \bibinfo
  {pages} {654} (\bibinfo {year} {1997}{\natexlab{b}})}\BibitemShut {NoStop}%
\bibitem [{\citenamefont {White}(1992)}]{White92}%
  \BibitemOpen
  \bibfield  {author} {\bibinfo {author} {\bibfnamefont {S.~R.}\ \bibnamefont
  {White}},\ }\bibfield  {title} {\bibinfo {title} {Density matrix formulation
  for quantum renormalization groups},\ }\href
  {https://doi.org/10.1103/PhysRevLett.69.2863} {\bibfield  {journal} {\bibinfo
   {journal} {Phys. Rev. Lett.}\ }\textbf {\bibinfo {volume} {69}},\ \bibinfo
  {pages} {2863} (\bibinfo {year} {1992})}\BibitemShut {NoStop}%
\bibitem [{\citenamefont {Eggert}\ and\ \citenamefont
  {Affleck}(1995)}]{Eggert95}%
  \BibitemOpen
  \bibfield  {author} {\bibinfo {author} {\bibfnamefont {S.}~\bibnamefont
  {Eggert}}\ and\ \bibinfo {author} {\bibfnamefont {I.}~\bibnamefont
  {Affleck}},\ }\bibfield  {title} {\bibinfo {title} {Impurities in
  $\mathit{S}\phantom{\rule{0ex}{0ex}}=\phantom{\rule{0ex}{0ex}}1/2$ heisenberg
  antiferromagnetic chains: Consequences for neutron scattering and knight
  shift},\ }\href {https://doi.org/10.1103/PhysRevLett.75.934} {\bibfield
  {journal} {\bibinfo  {journal} {Phys. Rev. Lett.}\ }\textbf {\bibinfo
  {volume} {75}},\ \bibinfo {pages} {934} (\bibinfo {year} {1995})}\BibitemShut
  {NoStop}%
\bibitem [{\citenamefont {Laukamp}\ \emph {et~al.}(1998)\citenamefont
  {Laukamp}, \citenamefont {Martins}, \citenamefont {Gazza}, \citenamefont
  {Malvezzi}, \citenamefont {Dagotto}, \citenamefont {Hansen}, \citenamefont
  {L\'opez},\ and\ \citenamefont {Riera}}]{Laukamp98}%
  \BibitemOpen
  \bibfield  {author} {\bibinfo {author} {\bibfnamefont {M.}~\bibnamefont
  {Laukamp}}, \bibinfo {author} {\bibfnamefont {G.~B.}\ \bibnamefont
  {Martins}}, \bibinfo {author} {\bibfnamefont {C.}~\bibnamefont {Gazza}},
  \bibinfo {author} {\bibfnamefont {A.~L.}\ \bibnamefont {Malvezzi}}, \bibinfo
  {author} {\bibfnamefont {E.}~\bibnamefont {Dagotto}}, \bibinfo {author}
  {\bibfnamefont {P.~M.}\ \bibnamefont {Hansen}}, \bibinfo {author}
  {\bibfnamefont {A.~C.}\ \bibnamefont {L\'opez}},\ and\ \bibinfo {author}
  {\bibfnamefont {J.}~\bibnamefont {Riera}},\ }\bibfield  {title} {\bibinfo
  {title} {Enhancement of antiferromagnetic correlations induced by nonmagnetic
  impurities: Origin and predictions for nmr experiments},\ }\href
  {https://doi.org/10.1103/PhysRevB.57.10755} {\bibfield  {journal} {\bibinfo
  {journal} {Phys. Rev. B}\ }\textbf {\bibinfo {volume} {57}},\ \bibinfo
  {pages} {10755} (\bibinfo {year} {1998})}\BibitemShut {NoStop}%
\bibitem [{\citenamefont {Frahm}\ and\ \citenamefont
  {Zvyagin}(1997)}]{Frahm97}%
  \BibitemOpen
  \bibfield  {author} {\bibinfo {author} {\bibfnamefont {H.}~\bibnamefont
  {Frahm}}\ and\ \bibinfo {author} {\bibfnamefont {A.~A.}\ \bibnamefont
  {Zvyagin}},\ }\bibfield  {title} {\bibinfo {title} {The open spin chain with
  impurity: an exact solution},\ }\href
  {https://doi.org/10.1088/0953-8984/9/45/021} {\bibfield  {journal} {\bibinfo
  {journal} {Journal of Physics: Condensed Matter}\ }\textbf {\bibinfo {volume}
  {9}},\ \bibinfo {pages} {9939} (\bibinfo {year} {1997})}\BibitemShut
  {NoStop}%
\bibitem [{\citenamefont {Affleck}\ \emph {et~al.}(1989)\citenamefont
  {Affleck}, \citenamefont {Gepner}, \citenamefont {Schulz},\ and\
  \citenamefont {Ziman}}]{Affleck1989}%
  \BibitemOpen
  \bibfield  {author} {\bibinfo {author} {\bibfnamefont {I.}~\bibnamefont
  {Affleck}}, \bibinfo {author} {\bibfnamefont {D.}~\bibnamefont {Gepner}},
  \bibinfo {author} {\bibfnamefont {H.~J.}\ \bibnamefont {Schulz}},\ and\
  \bibinfo {author} {\bibfnamefont {T.}~\bibnamefont {Ziman}},\ }\bibfield
  {title} {\bibinfo {title} {Critical behaviour of spin-s heisenberg
  antiferromagnetic chains: analytic and numerical results},\ }\href
  {https://doi.org/10.1088/0305-4470/22/5/015} {\bibfield  {journal} {\bibinfo
  {journal} {Journal of Physics A: Mathematical and General}\ }\textbf
  {\bibinfo {volume} {22}},\ \bibinfo {pages} {511} (\bibinfo {year}
  {1989})}\BibitemShut {NoStop}%
\bibitem [{\citenamefont {Singh}\ \emph {et~al.}(1989)\citenamefont {Singh},
  \citenamefont {Fisher},\ and\ \citenamefont {Shankar}}]{Singh1989}%
  \BibitemOpen
  \bibfield  {author} {\bibinfo {author} {\bibfnamefont {R.~R.~P.}\
  \bibnamefont {Singh}}, \bibinfo {author} {\bibfnamefont {M.~E.}\ \bibnamefont
  {Fisher}},\ and\ \bibinfo {author} {\bibfnamefont {R.}~\bibnamefont
  {Shankar}},\ }\bibfield  {title} {\bibinfo {title} {Spin-1/2
  antiferromagnetic xxz chain: New results and insights},\ }\href
  {https://doi.org/10.1103/PhysRevB.39.2562} {\bibfield  {journal} {\bibinfo
  {journal} {Phys. Rev. B}\ }\textbf {\bibinfo {volume} {39}},\ \bibinfo
  {pages} {2562} (\bibinfo {year} {1989})}\BibitemShut {NoStop}%
\bibitem [{\citenamefont {Hulth\'en}(1938)}]{Hulthen1938}%
  \BibitemOpen
  \bibfield  {author} {\bibinfo {author} {\bibfnamefont {L.}~\bibnamefont
  {Hulth\'en}},\ }\bibfield  {title} {\bibinfo {title} {Arkiv math. astron.
  fys. 26a},\ }\href@noop {} {\bibfield  {journal} {\bibinfo  {journal} {No11}\
  } (\bibinfo {year} {1938})}\BibitemShut {NoStop}%
\bibitem [{\citenamefont {Yang}\ and\ \citenamefont
  {Yang}(1966{\natexlab{a}})}]{Yang1966-1}%
  \BibitemOpen
  \bibfield  {author} {\bibinfo {author} {\bibfnamefont {C.~N.}\ \bibnamefont
  {Yang}}\ and\ \bibinfo {author} {\bibfnamefont {C.~P.}\ \bibnamefont
  {Yang}},\ }\bibfield  {title} {\bibinfo {title} {One-dimensional chain of
  anisotropic spin-spin interactions. i. proof of bethe's hypothesis for ground
  state in a finite system},\ }\href {https://doi.org/10.1103/PhysRev.150.321}
  {\bibfield  {journal} {\bibinfo  {journal} {Phys. Rev.}\ }\textbf {\bibinfo
  {volume} {150}},\ \bibinfo {pages} {321} (\bibinfo {year}
  {1966}{\natexlab{a}})}\BibitemShut {NoStop}%
\bibitem [{\citenamefont {Yang}\ and\ \citenamefont
  {Yang}(1966{\natexlab{b}})}]{Yang1966-2}%
  \BibitemOpen
  \bibfield  {author} {\bibinfo {author} {\bibfnamefont {C.~N.}\ \bibnamefont
  {Yang}}\ and\ \bibinfo {author} {\bibfnamefont {C.~P.}\ \bibnamefont
  {Yang}},\ }\bibfield  {title} {\bibinfo {title} {One-dimensional chain of
  anisotropic spin-spin interactions. ii. properties of the ground-state energy
  per lattice site for an infinite system},\ }\href
  {https://doi.org/10.1103/PhysRev.150.327} {\bibfield  {journal} {\bibinfo
  {journal} {Phys. Rev.}\ }\textbf {\bibinfo {volume} {150}},\ \bibinfo {pages}
  {327} (\bibinfo {year} {1966}{\natexlab{b}})}\BibitemShut {NoStop}%
\bibitem [{\citenamefont {Luther}\ and\ \citenamefont
  {Peschel}(1975)}]{Luther1975}%
  \BibitemOpen
  \bibfield  {author} {\bibinfo {author} {\bibfnamefont {A.}~\bibnamefont
  {Luther}}\ and\ \bibinfo {author} {\bibfnamefont {I.}~\bibnamefont
  {Peschel}},\ }\bibfield  {title} {\bibinfo {title} {Calculation of critical
  exponents in two dimensions from quantum field theory in one dimension},\
  }\href {https://doi.org/10.1103/PhysRevB.12.3908} {\bibfield  {journal}
  {\bibinfo  {journal} {Phys. Rev. B}\ }\textbf {\bibinfo {volume} {12}},\
  \bibinfo {pages} {3908} (\bibinfo {year} {1975})}\BibitemShut {NoStop}%
\bibitem [{\citenamefont {Bogoliubov}\ \emph {et~al.}(1986)\citenamefont
  {Bogoliubov}, \citenamefont {Izergin},\ and\ \citenamefont
  {Korepin}}]{Bogoliubov1986}%
  \BibitemOpen
  \bibfield  {author} {\bibinfo {author} {\bibfnamefont {N.}~\bibnamefont
  {Bogoliubov}}, \bibinfo {author} {\bibfnamefont {A.}~\bibnamefont
  {Izergin}},\ and\ \bibinfo {author} {\bibfnamefont {V.}~\bibnamefont
  {Korepin}},\ }\bibfield  {title} {\bibinfo {title} {Critical exponents for
  integrable models},\ }\href
  {https://doi.org/https://doi.org/10.1016/0550-3213(86)90579-1} {\bibfield
  {journal} {\bibinfo  {journal} {Nuclear Physics B}\ }\textbf {\bibinfo
  {volume} {275}},\ \bibinfo {pages} {687} (\bibinfo {year}
  {1986})}\BibitemShut {NoStop}%
\bibitem [{\citenamefont {Takigawa}\ \emph {et~al.}(1997)\citenamefont
  {Takigawa}, \citenamefont {Motoyama}, \citenamefont {Eisaki},\ and\
  \citenamefont {Uchida}}]{Takigawa97}%
  \BibitemOpen
  \bibfield  {author} {\bibinfo {author} {\bibfnamefont {M.}~\bibnamefont
  {Takigawa}}, \bibinfo {author} {\bibfnamefont {N.}~\bibnamefont {Motoyama}},
  \bibinfo {author} {\bibfnamefont {H.}~\bibnamefont {Eisaki}},\ and\ \bibinfo
  {author} {\bibfnamefont {S.}~\bibnamefont {Uchida}},\ }\bibfield  {title}
  {\bibinfo {title} {Field-induced staggered magnetization near impurities in
  the s= one-dimensional heisenberg antiferromagnet
  ${\mathrm{sr}}_{2}$${\mathrm{cuo}}_{3}$},\ }\href
  {https://doi.org/10.1103/PhysRevB.55.14129} {\bibfield  {journal} {\bibinfo
  {journal} {Phys. Rev. B}\ }\textbf {\bibinfo {volume} {55}},\ \bibinfo
  {pages} {14129} (\bibinfo {year} {1997})}\BibitemShut {NoStop}%
\bibitem [{\citenamefont {White}\ \emph {et~al.}(2002)\citenamefont {White},
  \citenamefont {Affleck},\ and\ \citenamefont {Scalapino}}]{White2002}%
  \BibitemOpen
  \bibfield  {author} {\bibinfo {author} {\bibfnamefont {S.~R.}\ \bibnamefont
  {White}}, \bibinfo {author} {\bibfnamefont {I.}~\bibnamefont {Affleck}},\
  and\ \bibinfo {author} {\bibfnamefont {D.~J.}\ \bibnamefont {Scalapino}},\
  }\bibfield  {title} {\bibinfo {title} {Friedel oscillations and charge
  density waves in chains and ladders},\ }\href
  {https://doi.org/10.1103/PhysRevB.65.165122} {\bibfield  {journal} {\bibinfo
  {journal} {Phys. Rev. B}\ }\textbf {\bibinfo {volume} {65}},\ \bibinfo
  {pages} {165122} (\bibinfo {year} {2002})}\BibitemShut {NoStop}%
\bibitem [{\citenamefont {Wessel}\ and\ \citenamefont
  {Haas}(2000)}]{Wessel2000}%
  \BibitemOpen
  \bibfield  {author} {\bibinfo {author} {\bibfnamefont {S.}~\bibnamefont
  {Wessel}}\ and\ \bibinfo {author} {\bibfnamefont {S.}~\bibnamefont {Haas}},\
  }\bibfield  {title} {\bibinfo {title} {Three-dimensional ordering in weakly
  coupled antiferromagnetic ladders and chains},\ }\href
  {https://doi.org/10.1103/PhysRevB.62.316} {\bibfield  {journal} {\bibinfo
  {journal} {Phys. Rev. B}\ }\textbf {\bibinfo {volume} {62}},\ \bibinfo
  {pages} {316} (\bibinfo {year} {2000})}\BibitemShut {NoStop}%
\bibitem [{\citenamefont {Des~Cloizeaux}\ and\ \citenamefont
  {Gaudin}(1966)}]{Des_Cloizeaux1966}%
  \BibitemOpen
  \bibfield  {author} {\bibinfo {author} {\bibfnamefont {J.}~\bibnamefont
  {Des~Cloizeaux}}\ and\ \bibinfo {author} {\bibfnamefont {M.}~\bibnamefont
  {Gaudin}},\ }\bibfield  {title} {\bibinfo {title} {Anisotropic linear
  magnetic chain},\ }\href {https://doi.org/10.1063/1.1705048} {\bibfield
  {journal} {\bibinfo  {journal} {Journal of Mathematical Physics}\ }\textbf
  {\bibinfo {volume} {7}},\ \bibinfo {pages} {1384} (\bibinfo {year} {1966})},\
  \Eprint {https://arxiv.org/abs/https://doi.org/10.1063/1.1705048}
  {https://doi.org/10.1063/1.1705048} \BibitemShut {NoStop}%
\bibitem [{\citenamefont {Kimura}\ \emph
  {et~al.}(2008{\natexlab{b}})\citenamefont {Kimura}, \citenamefont {Takeuchi},
  \citenamefont {Okunishi}, \citenamefont {Hagiwara}, \citenamefont {He},
  \citenamefont {Kindo}, \citenamefont {Taniyama},\ and\ \citenamefont
  {Itoh}}]{Kimura2008-2}%
  \BibitemOpen
  \bibfield  {author} {\bibinfo {author} {\bibfnamefont {S.}~\bibnamefont
  {Kimura}}, \bibinfo {author} {\bibfnamefont {T.}~\bibnamefont {Takeuchi}},
  \bibinfo {author} {\bibfnamefont {K.}~\bibnamefont {Okunishi}}, \bibinfo
  {author} {\bibfnamefont {M.}~\bibnamefont {Hagiwara}}, \bibinfo {author}
  {\bibfnamefont {Z.}~\bibnamefont {He}}, \bibinfo {author} {\bibfnamefont
  {K.}~\bibnamefont {Kindo}}, \bibinfo {author} {\bibfnamefont
  {T.}~\bibnamefont {Taniyama}},\ and\ \bibinfo {author} {\bibfnamefont
  {M.}~\bibnamefont {Itoh}},\ }\bibfield  {title} {\bibinfo {title} {Novel
  ordering of an $s=1/2$ quasi-1d ising-like antiferromagnet in magnetic
  field},\ }\href {https://doi.org/10.1103/PhysRevLett.100.057202} {\bibfield
  {journal} {\bibinfo  {journal} {Phys. Rev. Lett.}\ }\textbf {\bibinfo
  {volume} {100}},\ \bibinfo {pages} {057202} (\bibinfo {year}
  {2008}{\natexlab{b}})}\BibitemShut {NoStop}%
\bibitem [{\citenamefont {Johnston}\ \emph {et~al.}(2000)\citenamefont
  {Johnston}, \citenamefont {Kremer}, \citenamefont {Troyer}, \citenamefont
  {Wang}, \citenamefont {Kl\''umper}, \citenamefont {Bud'ko}, \citenamefont
  {Panchula},\ and\ \citenamefont {Canfield}}]{Johnston2000}%
  \BibitemOpen
  \bibfield  {author} {\bibinfo {author} {\bibfnamefont {D.~C.}\ \bibnamefont
  {Johnston}}, \bibinfo {author} {\bibfnamefont {R.~K.}\ \bibnamefont
  {Kremer}}, \bibinfo {author} {\bibfnamefont {M.}~\bibnamefont {Troyer}},
  \bibinfo {author} {\bibfnamefont {X.}~\bibnamefont {Wang}}, \bibinfo {author}
  {\bibfnamefont {A.}~\bibnamefont {Kl\''umper}}, \bibinfo {author}
  {\bibfnamefont {S.~L.}\ \bibnamefont {Bud'ko}}, \bibinfo {author}
  {\bibfnamefont {A.~F.}\ \bibnamefont {Panchula}},\ and\ \bibinfo {author}
  {\bibfnamefont {P.~C.}\ \bibnamefont {Canfield}},\ }\bibfield  {title}
  {\bibinfo {title} {Thermodynamics of spin $s=1/2$ antiferromagnetic uniform
  and alternating-exchange heisenberg chains},\ }\href
  {https://doi.org/10.1103/PhysRevB.61.9558} {\bibfield  {journal} {\bibinfo
  {journal} {Phys. Rev. B}\ }\textbf {\bibinfo {volume} {61}},\ \bibinfo
  {pages} {9558} (\bibinfo {year} {2000})}\BibitemShut {NoStop}%
\bibitem [{\citenamefont {Bera}\ \emph
  {et~al.}(2014{\natexlab{b}})\citenamefont {Bera}, \citenamefont {Lake},
  \citenamefont {Stein},\ and\ \citenamefont {Zander}}]{bera2014magnetic}%
  \BibitemOpen
  \bibfield  {author} {\bibinfo {author} {\bibfnamefont {A.}~\bibnamefont
  {Bera}}, \bibinfo {author} {\bibfnamefont {B.}~\bibnamefont {Lake}}, \bibinfo
  {author} {\bibfnamefont {W.-D.}\ \bibnamefont {Stein}},\ and\ \bibinfo
  {author} {\bibfnamefont {S.}~\bibnamefont {Zander}},\ }\bibfield  {title}
  {\bibinfo {title} {Magnetic correlations of the quasi-one-dimensional
  half-integer spin-chain antiferromagnets srm$_2$v$_2$o$_8$ (m= co, mn)},\
  }\href@noop {} {\bibfield  {journal} {\bibinfo  {journal} {Physical Review
  B}\ }\textbf {\bibinfo {volume} {89}},\ \bibinfo {pages} {094402} (\bibinfo
  {year} {2014}{\natexlab{b}})}\BibitemShut {NoStop}%
\end{thebibliography}%


%

\end{document}